# New Quantum Internet Applications via Verifiable One-Time Programs

**Lev Stambler**
levstamb@umd.edu
University of Maryland, College Park,
NeverLocal Ltd.

September 19, 2025

**Abstract**

We introduce Verifiable One-Time Programs (Ver-OTPs) and use them to construct single-round Open Secure Computation (OSC), a novel primitive enabling applications like (1) single-round sealed-bid auctions, (2) single-round and honest-majority atomic proposes—a building block of consensus protocols, (3) fair-exchange, and (4) single-round differentially private statistical aggregation without pre-registration. The underlying quantum requirement is minimal: only single-qubit states are needed alongside a hardware assumption on the receiver's quantum resources. Our work therefore provides a new framework for quantum-assisted cryptography that may be implementable with near-term quantum technology.

## 1 Introduction

The quantum internet promises many revolutionary applications beyond QKD and position verification, but most proposals require sophisticated quantum resources far beyond current technology. We demonstrate that simple BB84-like quantum states, already deployable over hundreds of meters, can enable novel quantum internet applications through our new primitive: verifiable one-time programs (Ver-OTPs) and our novel single-round open secure computation (OSC) framework.

We start by noticing that a flurry of recent constructions have built one-time programs (OTP s) from BB84-like states combined with various hardware assumptions [2, 3, 6, 10, 19–22, 29, 30].[1] The use of single-qubit states is particularly appealing as they are relatively easy to create, manipulate, and transmit over long distances compared to more complex quantum states. Unfortunately, coherence times of single-qubit states are typically short, and thus OTP s built from single-qubit states are inherently *ephemeral*. We call such OTPs *ephemeral one-time programs* (eOTPs).[2] At first glance, the utility of eOTPs seems limited. An eOTP that exists only for milliseconds would seem to require the receiver to be online and prepared for its immediate use, seemingly defeating the purpose of a "program" that can be evaluated at a later time.

[1]Hardware assumptions include classically-accessible oracles, bounded and noisy quantum storage, and limited quantum computation. Alternatively, restricting the class of functions computable by the OTP can also lead to constructions with weaker assumptions [15, 16].

[2]Surprisingly, ephemeral OTPs with a classical sender can construct powerful cryptographic primitives such as RAM obfuscation [31], though these constructions require heavy quantum resources due to their reliance on remote-state-preparation types of protocols.

However, recent experiments have demonstrated the feasibility of probabilistic one-time programs over significant distances (e.g., 650m of fiber [27]), suggesting that eOTPs are experimentally viable.

## 1.1 Our Contributions

We demonstrate that ephemeral OTPs, despite their limitations, are surprisingly powerful. Our contributions are twofold:
- We introduce ***verifiable one-time programs*** (Ver-OTPs), a new primitive that allows a receiver to verify that an OTP is well-formed *before* use with respect to some relation to publicly known data. This verification is non-interactive and reveals nothing about the secret program-specific data beyond its validity.
- Building on Ver-OTPs, we introduce ***open secure computation*** (OSC), a novel single-round secure computation model that requires no pre-registration.

Using OSC, we then show how to construct a variety of useful single-round protocols, including:
- Sealed-bid auctions,
- Honest-majority atomic proposals, a key component of consensus protocols,
- Differentially private statistical aggregation without any pre-registration,

We believe Ver-OTPs are of independent interest, with potential applications in fair exchange and pay-to-use programs.

In our constructions, we consciously trade classical computation for quantum simplicity (which faces fundamental physical limits).[3] The quantum component is minimal, requiring only single-qubit BB84-like states for the underlying OTPs, while the classical component makes use of multi-key homomorphic encryption (MHE), non-interactive zero-knowledge proofs (NIZKs), garbled circuits, and commitment schemes. We are thus optimistic that our protocols can be implemented with near-term quantum technology.

We do stress that our usage of multi-key homomorphic encryption is non-trivial and requires some modifications to existing schemes. We outline these modifications in Section 4 as well as some justification for why we believe that these modifications can be achieved with existing schemes. Still, we do not provide a full construction and leave this to future work.

## 1.2 Technical Overview

We now give a high-level overview of our techniques: first, we outline our construction of simulation-secure Ver-OTPs as they are one of the main building blocks of our OSC construction. We then outline how we formalize OSC, its construction, and its applications.

### 1.2.1 Verifiable One-Time Programs

Recall that a one-time program for a function $f(s, \cdot)$ allows for the evaluation of $f(s, x)$ for a single input $x$ of the receiver's choice. Generally, OTPs are constructed from a garbled circuit,

[3] We believe this trade-off is practical: classical computation is already highly optimized, improving rapidly over time: many classical cryptography works are already focused on optimizing the classical components of our protocols [18, 24, 26].

$\hat{f}(s, \cdot)$ and one-time memories ($\mathsf{OTM}$s) that provide the wire labels corresponding to a single input of the receiver's choice [13]. Our construction of *verifiable* $\mathsf{OTP}s$ (Ver-$\mathsf{OTP}$s) makes both components verifiable. We ensure that the garbled circuit is correct using a standard non-interactive zero-knowledge proof of knowledge ($\mathsf{NIZK}$) for correct garbling and some relation $\mathcal{R}$ on secret data $s$. The main novelty then lies in our construction of verifiable $\mathsf{OTM}$s.

**Verifiable One-Time Memories**

To construct verifiable $\mathsf{OTM}$s, we combine a cut-and-choose technique with secret sharing. For each of the two secret wire labels $(\kappa_0, \kappa_1)$, we generate $\zeta$ secret shares. These shares are then placed into $\zeta$ different, single-bit input $\mathsf{OTP}$s.[4] The $j$-th of these single-bit $\mathsf{OTP}$s has the following functionality for input bit $b$ in $\{0, 1\}$:

1. Output $\mathsf{sh}_{j,b}$ where $\mathsf{sh}_{j,b}$ is the $j$-th share of a secret sharing of $\kappa_b$ with a recovery threshold of $\frac{\zeta}{2} + 1$.
2. Output $\pi_{\mathsf{sh},j}$ which is a proof that $\mathsf{sh}_{j,b}$ is a valid share for the input wire label $\kappa_b$ of garbled circuit $\hat{f}(s, \cdot)$.

The receiver verifies correctness by randomly selecting a small fraction of these $\mathsf{OTP}$s to "open" (e.g. $\frac{\zeta}{8}$ for input 0 and $\frac{\zeta}{8}$ for input 1). Each opened $\mathsf{OTP}$ must provide a valid share and a proof of validity of the share. If all checks pass, the receiver is assured that, with overwhelming probability, a super-majority of the remaining, unopened $\mathsf{OTP}$s are correctly formed. Then, to recover the secret wire label $\kappa_b$ for input bit $b$, the receiver executes the unopened $\mathsf{OTP}$s for input $b$ and recover enough shares to reconstruct $\kappa_b$.

Critically, the secret sharing threshold is set to $\frac{\zeta}{2} + 1$, making it impossible for the receiver to recover *both* $\kappa_0$ and $\kappa_1$ as this would require opening more shares than are available.

**Technical Challenges**

Within our construction of Ver-$\mathsf{OTP}$s, we use UC-secure NIZKs, commitments, and $\mathsf{OTP}$s to simplify our proofs; but, we note that our proof requires rewinding the receiver's output to simulate which garbled circuit inputs are learned by the receiver. Thus, we only achieve simulation-based security rather than UC-security.

Further, we introduce a simple, global common reference string (CRS) with a trapdoor which allows the simulator to equivocate on the $\mathsf{NIZK}$ proof for correct garbling and relation check. We do not know of a way to remove the CRS assumption, though it does not contribute to the protocol's correctness. We leave the construction of Ver-$\mathsf{OTP}$s without a CRS as an open question.

### 1.2.2 Open Secure Computation (OSC)

Open secure computation ($\mathsf{OSC}$) is a new form of secure computation where no pre-registration or setup round is necessary and where a set of unknown sending parties can send inputs to a known (and untrusted) receiving party. The receiving party can then compute a function on the inputs of the sending parties. Within our construction of $\mathsf{OSC}$, we use only a *single round of communication.*

The use of a single round and lack of pre-registration simplifies the security proof and protocol but also introduces new edge cases which must be handled within our ideal specification.

[4] $\zeta = \zeta(\lambda)$ is a parameter that grows with the security parameter $\lambda$; the soundness error of the cut-and-choose is $\exp(-\Omega(\zeta))$.

Specifically, we define the ideal functionality of OSC roughly as follows for some function $f : (\{0,1\}^a \cup \{\perp\})^k \to \{0,1\}^c$ where $k$ is the maximum number of inputs from the sending parties for $f$:

1. The sending parties send their inputs to the ideal functionality.
2. The receiving party is then allowed to adaptively "partition" the sent inputs into a number of disjoint groups, $\mathcal{S}_1, ..., \mathcal{S}_p$.[5] Abusing our notation, the receiving party can then compute $f(\mathcal{S}_1), ..., f(\mathcal{S}_p)$ on the inputs of each group.[6]

Though this formulation may seem odd, we believe it to be necessary as an untrusted receiving party may always choose to ignore some of the sending parties in one of the function computations and include them in another function computation. As we outline in Section 6, our formulation of OSC can be made immune to partitioning attacks if we have some sort of "pre-registration" of the sending parties and an honest majority requirement. Specifically, we can prevent partitioning by having the function $f$ check that the inputs are from the registered parties and that a majority of inputs are present. Then, the receiver cannot partition the inputs as any partition which does not contain a majority of the registered parties will output $\perp$.

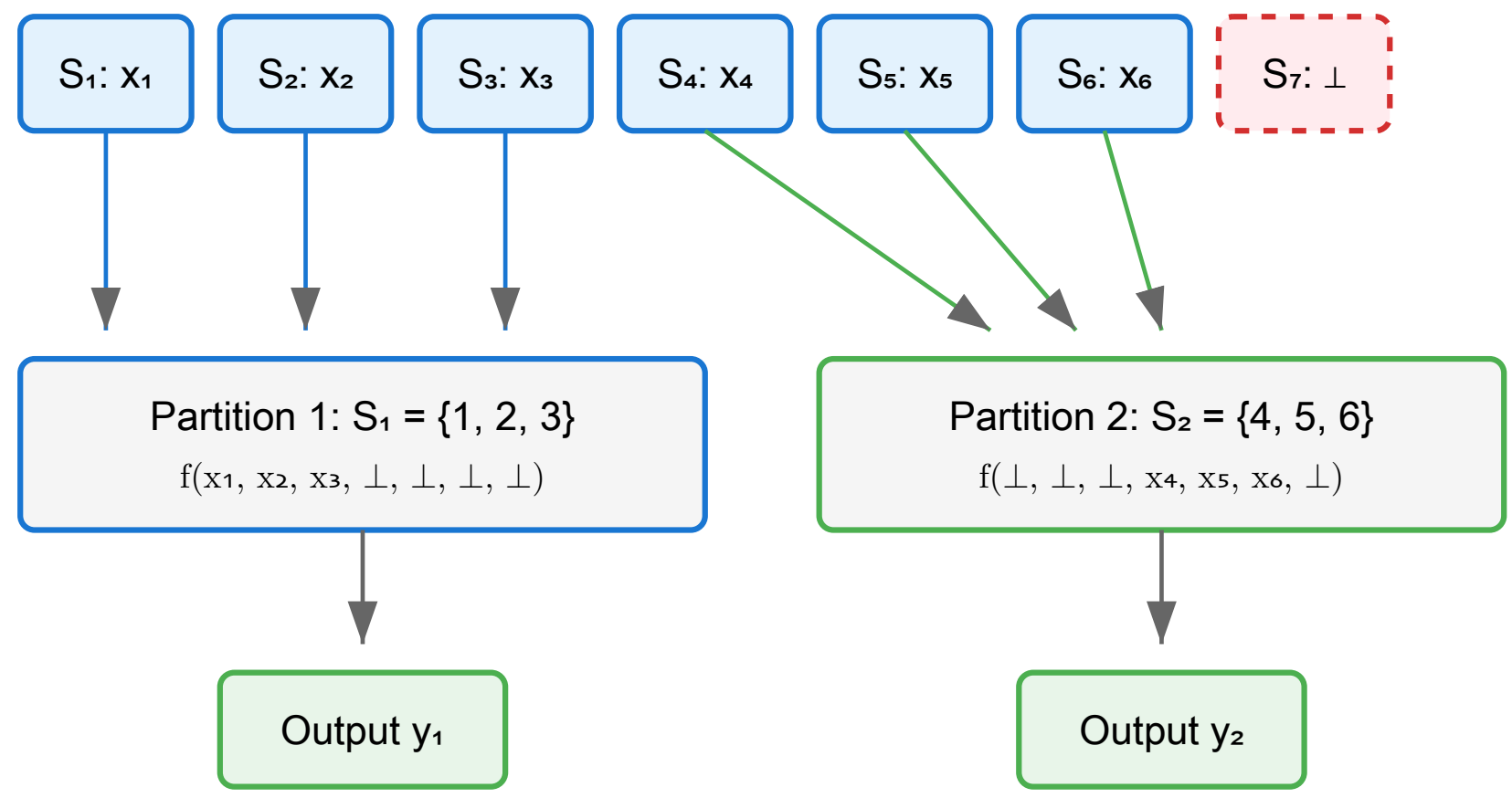

Figure 1: OSC functionality for function $f$ with sending parties $P_1, ..., P_7$ and receiving party which makes two partitions, $\mathcal{S}_1 = \{1,2,3\}$ and $\mathcal{S}_2 = \{4,5,6\}$. $P_7$ is corrupted and thus its input is set to $\perp$ in the protocol.

Prior to going over the construction of OSC, we briefly outline some applications of OSC to motivate our definition:

- **Single round sealed bid auctions**: In a sealed bid auction, a set of bidders wish to submit bids to an auctioneer who then determines the winner of the auction and the price they must pay. We can use OSC with pre-registration to construct a sealed bid auction where the auctioneer can only run the auction once and must include a majority of the registered bidders in the auction. Then the function, $f$, would take in the bids of all registered bidders and output the winner, the price they must pay, and any other secret information to execute the payment (i.e. a transaction signature for a cryptocurrency payment).
- **Honest majority, single round atomic proposals**: In an honest majority atomic proposal, a leader wishes to propose a value and get a set of attestations from a majority of the parties. This is a key component of many consensus protocols such as PBFT [8], HotStuff [33], and Ethereum's consensus protocol [12]. We can use OSC with pre-registration to construct atomic proposals where a chosen leader acts as the receiving party and the other parties act as sending parties. Each sending party then sends the receiver input which allows the receiver to compute a signature on the agreed upon value if and only if a majority of the registered parties agree on the value and sent their input to the receiver. Thus, the receiver can only run the protocol once and must include a majority of the registered parties in the protocol.

[5] Here, "adaptivity" means that the receiving party can choose the first partition, get the output of $f$ on that partition, and then choose the next partition based on the output of the previous partition. The receiving party can do this until all inputs have been partitioned or the receiving party decides to stop.

[6] If $|\mathcal{S}_i| < k$, then the receiving party is allowed to pad the input with $\perp$ values to make the input size $k$.

- **Single round statistical aggregation**: In a statistical aggregation protocol, a set of parties wish to compute some statistic on their private data without revealing their private data. Using a *differential privacy* mechanism [11], we can ensure that the output of the statistical aggregation does not leak too much information about any single party's input. We can use OSC *without* any pre-registration to construct such a protocol by having each party submit their data to be aggregated and some secret value which is used to seed the noise generation of the differential privacy mechanism. Then, the function $f$ would compute the desired statistic and use the secret values to generate a single noise value to add to the statistic.

### Construction of OSC

Our construction of OSC makes use of multi-key homomorphic encryption (MHE) and verifiable one-time programs (Ver-OTPs). A multi-key homomorphic encryption scheme allows for multiple parties to generate public/secret key pairs and encrypt messages under their public keys [1, 23]. Then, given a set of ciphertexts encrypted under different public keys, it is possible to homomorphically evaluate a function on the underlying plaintexts and produce a "joint" ciphertext. This joint ciphertext can then be partially decrypted by each party using their secret key and then fully decrypted given all partial decryptions.

Roughly, the protocol works as follows:

- Each sending party generates a public/secret key pair, $\mathtt{pk}_i, \mathtt{sk}_i$, for the MHE scheme and sends their public key to the receiving party.
- Each sending party then encrypts their input under their public key and sends the ciphertext to the receiving party.
- Each sending party also creates a Ver-OTP for the function $g(\mathtt{sk}_i, \mathtt{ct}_1, ..., \mathtt{ct}_k)$ which is hardcoded to use the sending party's secret key to partially decrypt the output of function $f$ evaluated on the ciphertexts $\mathtt{ct}_1, ..., \mathtt{ct}_k$. The Ver-OTP will only pass verification if the embedded secret key $\mathtt{sk}_i$ is the secret key corresponding to the inputted ciphertext $\mathtt{ct}_i$. The sending party then sends this Ver-OTP to the receiving party as well.
- The receiving party then verifies the Ver-OTPs. If any proofs of the Ver-OTPs fail, the receiving party then replaces the corresponding ciphertext with an encryption of $\perp$ under their own MHE public key. Then, the receiving party can choose to partition the sending parties into a number of disjoint groups, $\mathcal{S}_1, ..., \mathcal{S}_p$.[7] For each group $\mathcal{S}_i$, the receiving party then homomorphically evaluates $f$ on the ciphertexts of the parties in $\mathcal{S}_i$ (and encryptions of $\perp$ for any missing parties) to produce a joint ciphertext $\widehat{\mathtt{ct}_{\mathcal{S}_i}}$. Then, the receiving party evaluates each Ver-OTP on the ciphertexts $(\mathtt{ct}_\ell)_{\ell \in \mathcal{S}_i}$ to produce a set of partial decryptions which can then be combined to produce the output of $f$ on the inputs of the parties in $\mathcal{S}_i$.

[7] In the actual scheme, the partitions can be chosen adaptively as the receiving party learns the output of each partition. For simplicity, we ignore this adaptive choice in our overview.

The security of the protocol follows from the security of the two underlying primitives, though we have to strengthen the security of MHE to allow for the following:

1. auxiliary inputs to the adversary.
2. the removal index dependence in key generation (which is already considered in some works [17]).
3. the ability to compute and partially decrypt on subsets of ciphertexts and keys: i.e. some computations may only involve a subset of the party's inputs.
4. the ability to simulate partial decryptions when not all partial decryptions are known to the adversary. Mainly, this simulation does not require knowledge of the function output.

We believe that these modifications can be achieved with existing schemes and provide some justification for our belief in Section 4. Though, we leave the construction of such a scheme to future work. Alternatively, we can also aim to modify our OSC construction to avoid the need for these modifications in future work. We give an updated full security definition in Section 4.

The key idea is that the MHE scheme ensures that the receiving party learns nothing about the inputs of the sending parties beyond that of the Ver-OTPs' output. The verifiability of the OTPs ensure that no sending party can cause the protocol to abort as we can always *check* that the Ver-OTPs are well-formed and will output a valid partial decryption.

## Paper Organization

In Section 2, we outline definitions and primitives which we use throughout the paper. Then, in Section 3, we outline both our definition and construction of Ver-OTPs. Following this, in Section 4, we provide the definitions for multi-key homomorphic encryption which we use in our OSC construction. In Section 5, we outline both our definition and construction of OSC. Our constructions are then employed in Section 6 to construct a number of useful applications. Finally, we conclude in Section 7 with a summary of our contributions and open questions for future work.

# 2 Preliminaries

We now define the cryptographic primitives we will be using in this work. We mainly use UC-secure primitives as they compose nicely in parallel [7].

We use simulation-based security definitions throughout. For a protocol $\Pi$ and adversary $\mathcal{A}$ corrupting a set of parties, $\mathsf{Real}_{\Pi,\mathcal{A}}(\cdot)$ denotes the joint output distribution of all parties in the real execution of $\Pi$. For an ideal functionality $\mathcal{F}$ and simulator $\mathsf{Sim}$, $\mathsf{Ideal}_{\mathcal{F},\,\mathsf{Sim}}(\cdot)$ denotes the joint output distribution in the ideal execution where honest parties interact with $\mathcal{F}$ and $\mathsf{Sim}$ simulates the corrupted parties. A protocol $\Pi$ securely realizes $\mathcal{F}$ if for every PPT (or QPT, in the quantum setting) adversary $\mathcal{A}$ there exists a PPT (resp.
QPT) simulator $\mathsf{Sim}$ such that $\mathsf{Real}_{\Pi,\mathcal{A}}(\cdot) \overset{c}{\approx} \mathsf{Ideal}_{\mathcal{F},\,\mathsf{Sim}}(\cdot)$.

## 2.1 Notation and Conventions

We let $\lambda$ denote the security parameter; all other quantities are implicitly polynomial in $\lambda$ unless stated otherwise. We write $\mathsf{negl}(\lambda)$ for any function negligible in $\lambda$. For our OTP and Ver-OTP constructions, $a, b, c$ denote the bit-lengths of the sender's secret input, the receiver's input, and the output respectively, so the function being computed is $f : \{0,1\}^a \times \{0,1\}^b \to \{0,1\}^c$.

The parameter $\zeta = \zeta(\lambda)$ controls the number of secret shares in the cut-and-choose verification of our Ver-OTP construction; it grows with $\lambda$ and the soundness error is $\exp(-\Omega(\zeta))$.

## 2.2 Non-Interactive Zero-Knowledge Arguments of Knowledge

Non-Interactive Zero-Knowledge Arguments (which we write as NIZK-Arg or NIZK) is a non-interactive cryptographic primitive that allows a prover to convince a verifier that they know a witness for a given statement without revealing any information about the witness itself beyond the validity of the statement. They are achieved either through the use of a common reference string (CRS) or the random oracle model [9, 14].

**Functionality $\mathcal{F}_{\mathsf{NIZK\text{-}Arg}}$**

Parameterized with relation $\mathcal{R}$, running with a prover $P$ and verifier $V$, and adversary $\mathcal{S}$:

**Proving Phase**: Upon receiving $(\textbf{prove}, \text{sid}, x, w)$ from $P$ ignore if $(x, w) \notin \mathcal{R}$. Send $(\textbf{prove}, x)$ to $\mathcal{S}$ and wait for answer $(\textbf{proof}, \pi)$. Upon receiving the answer store $(x, \pi)$ and send $(\textbf{proof}, \text{sid}, \pi)$ to $P$.

**Verifying Phase**: Upon receiving $(\textbf{verify}, \text{sid}, x, \pi)$ from party $V$ check whether $(x, \pi)$ is stored. If not, send $(\textbf{verify}, x)$ to $\mathcal{S}$ and wait for answer $(\textbf{witness}, w)$. Upon receiving the answer, check whether $(x, w) \in \mathcal{R}$ and if yes, store $(x, \pi)$. If $(x, \pi)$ has been stored return $(\textbf{verification}, \text{sid}, \texttt{accept})$ to $V$ else return $(\textbf{verification}, \text{sid}, \texttt{reject})$.

Ideal Functionality 1: NIZK argument ideal functionality $\mathcal{F}_{\mathsf{NIZK\text{-}Arg}}$ [14]

## 2.3 One-Time Programs (OTP)

Though impossible in the standard model of classical and quantum computation [6, 13], one-time programs can be instantiated using a variety of hardware assumptions such as one-time memories (OTM) [13] or using quantum states [3, 6, 19, 21, 29, 30].

**Functionality $\mathcal{F}_{\mathsf{OTP}}$**

Running with a sender $S$ and receiver $R$, and adversary $\mathcal{S}$:

**Create**: Upon receiving $(\textbf{create}, \text{sid}, s_{\mathsf{otp}})$ from the sender, send **create** to the receiver and store $s_{\mathsf{otp}}$

**Execute**: Upon receiving $(\textbf{execute}, \text{sid}, x \in \{0,1\}^b)$ from the receiver, if $s_{\mathsf{otp}}$ is stored, compute $y = f(s_{\mathsf{otp}}, x)$, send $y$ to the receiver, and delete $s_{\mathsf{otp}}$.

Ideal Functionality 2: OTP functionality $\mathcal{F}_{\mathsf{OTP}}$ for poly-time $f : \{0,1\}^a \times \{0,1\}^b \to \{0,1\}^c$ [6]

## 2.4 Commitments

Commitment schemes are cryptographic primitives that allow one party to commit to a chosen value while keeping it hidden from others, with the ability to reveal the committed value later.

**Functionality $\mathcal{F}_{\mathsf{Comm}}$**

Running parties $P_1, ..., P_n$ and adversary $\mathcal{S}$ for message space $\mathcal{M}$:

**Commit Phase**: Upon receiving $(\textbf{commit}, \text{sid}, \textbf{cid}, P_i, P_j, m \in \mathcal{M})$ from committer $P_i$, record $(\textbf{cid}, P_i, P_j, m)$ and send $(\textbf{receipt}, \text{sid}, \textbf{cid}, P_i, P_j)$ to $P_j$ and $\mathcal{S}$. Ignore any subsequent $(\textbf{commit}, \text{sid}, \textbf{cid}, P_i, P_j, m')$ from $P_i$ for any $m' \in \mathcal{M}$.

**Opening Phase**: Upon receiving $(\textbf{open}, \text{sid}, \textbf{cid}, P_i, P_j)$ from committer $P_i$ proceed as follows: if $(\textbf{cid}, P_i, P_j, m)$ is recorded, send $(\textbf{open}, \text{sid}, \textbf{cid}, P_i, P_j, m)$ to $P_j$ and $\mathcal{S}$. Otherwise, do nothing.

Ideal Functionality 3: Commitment ideal functionality $\mathcal{F}_{\mathsf{Comm}}$ [7]

## 2.5 Garbled Circuits

We use garbled circuits as defined by Ref. [4] though we modify the definition to allow for multi-bit outputs.

**Definition 2.1** *(Garbled Circuit, [4, 32])*: Let $\{\mathcal{C}_n\}_n$ be a family of circuits where each circuit in $\mathcal{C}_n$ has $n$ bit inputs. A garbling scheme, $\mathsf{GC}$, for circuit family $\{\mathcal{C}_n\}_n$ consists of polynomial-time algorithms $(\mathsf{Garble}, \textbf{Eval})$:

- $\mathsf{Garble}(1^\lambda, C \in \mathcal{C}_n)$: The algorithm takes a security parameter $\lambda$ and a circuit $C$ as input and outputs a garbled circuit $G$ together with $2n$ wire keys $\{w_{i,b}\}_{i \in [n], b \in \{0,1\}}$
- $\textbf{Eval}\left(G, \{w_i\}_{i \in [n]}\right)$: The evaluation algorithm takes as input a garbled circuit $G$ and $n$ wire keys $\{w_i\}_{i \in [n]}$ and outputs $y \in \{0,1\}^m$.

**Definition 2.2** *(Garbling Correctness)*: A garbling scheme $\mathsf{GC}$ for circuit family $\{\mathcal{C}_n\}_n$ is said to be correct if for all $\lambda$, $n$, $x \in \{0,1\}^n$ and $C \in \mathcal{C}_n$,

$$\textbf{Eval}\left(G, \left\{w_{i,x_i}\right\}(i \in [n])\right) = C(x),$$

where

$$\left(G, \left\{w_{i,b}\right\}_{i \in [n], b \in \{0,1\}}\right) \leftarrow \mathsf{Garble}(1^\lambda, C).$$

**Definition 2.3** *(Selective Security)*: A garbling scheme $\mathsf{GC} = (\mathsf{Garble}, \textbf{Eval})$ for a class of circuits $\mathcal{C} = \{\mathcal{C}_n\}_n$ is said to be a selectively secure garbling scheme if there exists a polynomial-time simulator Sim such that for all $\lambda$, $n$, $C \in \mathcal{C}_n$ and $x \in \{0,1\}^n$, the following holds:

$$\left\{\text{Sim}\left(1^\lambda, 1^n, 1^{|C|}, C(x)\right)\right\} \overset{c}{\approx} \left\{\left(G, \left\{w_{i,x_i}\right\}_{i \in [n]}\right) : \left(G, \left\{w_{i,b}\right\}_{i \in [n], b \in \{0,1\}}\right) \leftarrow \mathsf{Garble}(1^\lambda, C)\right\}$$

## 2.6 Secret Sharing

We also use information-theoretic secret sharing schemes [5, 28]. We will use $(t, n)$-secret sharing schemes where a secret is divided into $n$ shares such that any $t$ shares can reconstruct the secret, but any set of $t-1$ or fewer shares reveals no information about the secret. Moreover, we will have it be "deterministic" in the sense that we will input our randomness to the sharing algorithm and thus the shares will be fixed for a given secret and randomness.

**Definition 2.4** *(Deterministic Secret Sharing Scheme)*: A *deterministic secret sharing scheme* consists of two algorithms, Share and Rec, such that:

- $\mathsf{Share}(i, s, n, t, r) \to s_i$: the sharing algorithm takes as input a party index $i \in \{1, ..., n\}$, a secret $s$, the total number of shares $n$, the threshold $t$, and randomness $r$ and outputs the share $s_i$ for party $i$.
- $\mathsf{Rec}\Big((s_i)_{\{i \in I\}}\Big) \to s$ for any set $I \subseteq \{1, ..., n\}$ with $|I| \geq t$: The reconstruction algorithm takes as input any set of $t$ shares and outputs the original secret $s$.

We then define soundness as follows:

**Definition 2.5** *(Soundness of Deterministic Secret Sharing Scheme)*: A deterministic secret sharing scheme is *sound* if for any two secrets $s, s'$ and uniform randomness $r$, then for any set $J \subseteq \{1, ..., n\}$ with $|J| \leq t - 1$, $s$ is information-theoretically uniformly random.

# 3 Verifiable One-Time Programs

We now introduce our new primitive, verifiable one-time programs (Ver-OTP), and then construct it from one-time programs, non-interactive zero-knowledge arguments of knowledge, commitment schemes, garbled circuits, and secret sharing schemes.

## 3.1 Ideal Functionality

We define the ideal functionality of verifiable one-time programs as follows in the simulation-based setting. We note that our definition allows for corruptions of either the sender or receiver. A corrupted sender can choose to set $s_{\mathsf{otp}}, z$ to any value, but if the receiver accepts, then it is guaranteed that $\mathcal{R}(s_{\mathsf{otp}}, z) = \mathtt{accept}$ for the maliciously chosen input $s_{\mathsf{otp}}$.

**Functionality $\mathcal{F}_{\text{Ver-}\mathsf{OTP}}$**

Parameterized with relation $\mathcal{R}$, running with a sender $S$ and a receiver $R$, and adversary $\mathcal{S}$. For security purposes, we also generate a re-usable common reference string `crs` and a trapdoor $\tau$ such that `crs` is a commitment to a random value and $\tau$ is the opening of the commitment. We write $z$ for the public input to the relation and $s_{\mathsf{otp}}$ for the secret input to the relation, which is also the secret input to the program $f$.

**Create**: Upon receiving $\left(\mathbf{create}, s_{\mathsf{otp}}, z, \mathtt{crs}\right)$ from the sender $S$:

- Ignore if $\left(z, s_{\mathsf{otp}}\right) \notin \mathcal{R}$
- Send $(\mathbf{prove}, z)$ to $\mathcal{S}$ and wait for answer $(\mathbf{proof}, s^*, z^*)$
- If $s^* \neq \perp$, set $s_{\mathsf{otp}} = s^*$. Set $z = z^*$
- If $\left(z, s_{\mathsf{otp}}\right) \notin \mathcal{R}$ ignore
- Otherwise, send $(\mathbf{create}, \mathsf{v\text{-}otp})$ to the sender, store $\left(z, s_{\mathsf{otp}}\right)$

**Verifying Phase**: Upon receiving $(\mathbf{verify}, \mathsf{v\text{-}otp}, \mathrm{z}, \mathtt{crs})$ from the receiver $R$:

- Check whether a pair $\left(z, s_{\mathsf{otp}}\right)$ was stored during the Create phase
- If $\left(z, s_{\mathsf{otp}}\right)$ has been stored return $(\mathbf{verification}, \mathsf{otp}, \mathtt{accept})$ to $R$ and store `accepted`
- Otherwise return $(\mathbf{verification}, \mathtt{reject})$ to $R$.

**Execute**: Upon receiving $\left(\mathbf{execute}, \mathsf{otp}, x \in \{0,1\}^b\right)$ from the receiver $R$, if $s_{\mathsf{otp}}$ is stored and `accepted` is stored, compute $y = f\left(s_{\mathsf{otp}}, x\right)$, send $y$ to $R$, and delete $s_{\mathsf{otp}}$ and `accepted`.

Ideal Functionality 4: Verifiable OTP ideal functionality $\mathcal{F}_{\text{Ver-}\mathsf{OTP}}$ for program $f : \{0,1\}^a \times \{0,1\}^b \to \{0,1\}^c$

**Definition 3.1** *(Simulation-Sound Verifiable One-Time Programs)*: We say that a protocol, $\Pi_{\text{Ver-}\mathsf{OTP}}$, is a simulation-sound verifiable one-time program if there exists a pair of BQP/ BPP simulators $\mathsf{Sim}_S, \mathsf{Sim}_R$ for the sender and receiver respectively such that the following holds, where $\mathsf{Real}_{\Pi_{\text{Ver-}\mathsf{OTP}},\mathcal{A},\text{ Party}}(\cdot)$ denotes the joint output of all parties in a real execution of $\Pi_{\text{Ver-}\mathsf{OTP}}$ with adversary $\mathcal{A}$ corrupting the indicated party, and $\mathsf{Ideal}_{\mathcal{F}_{\text{Ver-}\mathsf{OTP}},\text{ }\mathsf{Sim},\text{ Party}}(\cdot)$ denotes the joint output in the corresponding ideal execution with $\mathcal{F}_{\text{Ver-}\mathsf{OTP}}$ and simulator $\mathsf{Sim}$:

$$\mathsf{Real}_{\Pi_{\text{Ver-}\mathsf{OTP}},\mathcal{A}(\mathcal{R},f,z,\mathsf{aux}),\text{ Sender}}\left(1^\lambda, s_{\mathsf{otp}}, \mathsf{aux}, z, f, \mathcal{R}\right) \overset{c}{\approx} \mathsf{Ideal}_{\mathcal{F}_{\text{Ver-}\mathsf{OTP}},\mathsf{Sim}_S,\text{ Sender}}\left(1^\lambda, s_{\mathsf{otp}}, \mathsf{aux}, z, f, \mathcal{R}\right)$$

and

$$\mathsf{Real}_{\Pi_{\text{Ver-}\mathsf{OTP}},\mathcal{A}(\mathcal{R},f,z,\mathsf{aux}),\text{ Receiver}}\left(1^\lambda, s_{\mathsf{otp}}, \mathsf{aux}, z, f, \mathcal{R}\right) \overset{c}{\approx} \mathsf{Ideal}_{\mathcal{F}_{\text{Ver-}\mathsf{OTP}},\mathsf{Sim}_R(\tau),\text{ Receiver}}\left(1^\lambda, s_{\mathsf{otp}}, \mathsf{aux}, z, f, \mathcal{R}\right)$$

where $\mathsf{aux}$ is independent of $s_{\mathsf{otp}}$.

## 3.2 Construction

We will first outline the relation and circuits used and then give our construction in Construction 4.

### Relations and Circuits

We now outline the relations and circuits used in our construction.

**Relation 1:** Modified relation $\hat{\mathcal{R}}$ for garbled circuit $\hat{f}$

$\hat{\mathcal{R}}$ for garbled circuit $\hat{f}$ with randomness $r$ be the relation:

$$\hat{\mathcal{R}}\left(\left(s_{\mathsf{otp}}, r, \left(\kappa_{i,\beta}\right)_{i,\beta}, \tau\right), \left(\hat{f}, z, \left(\mathtt{comm}_{i,\beta}\right)_{i\in[b],\beta\in\{0,1\}}, \mathtt{crs}\right)\right) = \mathtt{accept}$$

if and only if all the following hold:

1. $\mathcal{F}_{\mathsf{Comm}}$ has a record $\mathtt{comm}_{i,\beta} = \left(\mathbf{cid}_{i,\beta}, S, R, \kappa_{i,\beta}\right)$
2. $\hat{f}$ is a valid garbling of $f\left(s_{\mathsf{otp}}, \cdot\right)$ with randomness $r$ and input wires labeled $\left(\kappa_{i,\beta}\right)_{i\in[b],\beta\in\{0,1\}}$
3. $\mathcal{R}\left(s_{\mathsf{otp}}, z\right) = \mathtt{accept}$

or $\tau$ is the opening of $\mathtt{crs}$

**Relation 2:** Per-wire Relation $\mathcal{R}_{i,\beta}$

$\mathcal{R}_{i,\beta}$ for $i \in [b]$ and $\beta \in \{0,1\}$ to be the relation such that

$$\mathcal{R}_{i,\beta}\left(\left(\kappa_{i,\beta}, r_{i,\beta}, \theta_{i,\beta}\right), \left(\mathsf{sh}, \alpha, \mathtt{comm}_{i,\beta}, \mathtt{comm}'_{i,\beta}\right)\right) = \mathtt{accept}$$

if and only if all the following hold:

$\mathcal{F}_{\mathsf{Comm}}$ has a record $\mathtt{comm}_{i,\beta} = \left(\mathbf{cid}_{i,\beta}, S, R, \kappa_{i,\beta}\right)$
$\mathcal{F}_{\mathsf{Comm}}$ has a record $\mathtt{comm}'_{i,\beta} = \left(\mathbf{cid}'_{i,\beta}, S, R, r_{i,\beta}\right)$
$\mathsf{sh}$ equals the $\alpha$-th element of $\mathsf{sh} \leftarrow \mathsf{Share}\left(\alpha, \kappa_{i,\beta}, \zeta, \frac{\zeta}{2} + 1, r_{i,\beta}\right)$

### 3.2.1 Construction of Ver-OTP

We now provide an additional circuit which will be used in our main construction.

**Circuit 3:** One-time memory program $f_{i,\alpha} : \{0,1\} \to \{0,1\}^{\lambda}$ for $\alpha \in [\zeta]$

**Hard-Coded:** Wires $\kappa_{i,0}, \kappa_{i,1}$, randomness $r_{i,0}, r_{i,1}$,
and for both $\beta \in \{0,1\}$:
$\mathsf{sh}^{\alpha}_{i,0} \leftarrow \mathsf{Share}\left(\alpha, \kappa_{i,\beta}, \zeta, \frac{\zeta}{2} + 1, r_{i,\beta}\right)$
$\pi^{\alpha}_{i,0}$ is response from $\left(\mathbf{prove}, \mathrm{sid} = \mathsf{NIZK}_{i,\beta}, \left(\kappa_{i,\beta}, r_{i,\beta}\right), \left(\mathsf{sh}^{\alpha}_{i,\beta}, \alpha, \mathtt{comm}_{i,\beta}, \mathtt{comm}_{i,\beta'}\right)\right)$ to $\mathcal{F}.\mathsf{NIZK\text{-}Arg}$

**On input** $\beta \in \{0,1\}$, return $\left(\mathsf{sh}^{\alpha}_{i,\beta}, \pi^{\alpha}_{i,\beta}\right)$

We now give the construction of our Ver-OTP in Construction 4. Given that we use a multi-commitment ideal functionality, we will let the session identifier, sid, for each commitment simply be $\mathrm{sid}_c = \mathrm{Commit}_1$.

**Theorem 3.1** (Security of Verifiable One-Time Programs): Assuming the existence of secure implementations of ideal functionalities $\mathcal{F}_{\mathsf{NIZK\text{-}Arg}}, \mathcal{F}_{\mathsf{Comm}}$, then Construction 4 is simulation secure for Ideal Functionality 4.

*Proof*: We proceed by first considering a malicious sender and then a malicious receiver in a similar manner to Broadbent et al. [6].

**Construction 4:** Ver-OTP with program $f$, relation $\mathcal{R}$, sending party $S$, receiving party $R$, and common reference string $\texttt{crs}$

Ver-$\mathsf{OTP}$.**create**$(s_{\mathsf{otp}}, z, \texttt{crs})$ for $s_{\mathsf{otp}} \in \{0,1\}^a$:

1. $\hat{f}, (\kappa_{i,\beta})_{i,\beta} \leftarrow \mathsf{Garble}(f(s_{\mathsf{otp}}, \cdot), r)$
2. **For** $i \in [b], \beta \in \{0,1\}$:
3.     Set $\mathbf{cid}_{i,\beta} \leftarrow (i, \beta, \mathrm{sid} = \mathrm{Commit}_1)$
4.     Send $(\mathbf{commit}, \mathrm{sid}_c, \mathbf{cid}_{i,\beta}, S, R, \kappa_{i,\beta})$ // Commit to garbled input wire labels
5.     Receive $\texttt{comm}_{i,\beta} = (\mathbf{receipt}, \mathrm{sid}_c, \mathbf{cid}_{i,\beta}, S, R)$
6. Send $\left(\mathbf{prove}, \mathrm{sid} = \mathsf{NIZK}_1, \left(s_{\mathsf{otp}}, (\kappa_{i,\beta})_{i,\beta}\right), \left(\hat{f}, z, (\texttt{comm}_{i,\beta})_{i,\beta}\right)\right)$ to $\mathcal{F}.\mathsf{NIZK}\text{-}\mathsf{Arg}_{\hat{\mathcal{R}}}$ and receive $(\mathbf{proof}, \mathrm{sid} = \mathsf{NIZK}_1, \pi)$ // Proof of garbling and relation satisfaction
7. **For** $i \in [b]$:
8.     $r_{i,\beta} \leftarrow \$$
9.     Send $(\mathbf{commit}, \mathrm{sid}_c, \mathbf{cid}'_{i,\beta}, S, R, r_{i,\beta})$ // Commit to randomness for secret sharing
10.     Receive $\texttt{comm}'_{i,\beta} = (\mathbf{receipt}, \mathrm{sid}_c, \mathbf{cid}'_{i,\beta}, S, R)$
11.     $\mathsf{otp}_{i,\alpha} = \mathsf{OTP}.\mathbf{Gen}(1^\lambda, f_{i,\alpha})$ for $\alpha \in [\zeta]$ // Generate OTPs for each input wire
12. **Return** $\mathsf{v\text{-}otp} = \left((\mathsf{otp}_{i,\alpha})_{i \in [b], \alpha \in [\zeta]}, (\texttt{comm}_{i,\beta})_{i,\beta}, (\texttt{comm}'_{i,\beta})_{i,\beta}, \hat{f}, \pi\right)$

Ver-$\mathsf{OTP}$.**Verify**$(\mathsf{v\text{-}otp}, z, \texttt{crs})$:

13. Parse $\mathsf{v\text{-}otp}$ as $\left((\mathsf{otp}_{i,\alpha})_{i \in [b], \alpha \in [\zeta]}, (\texttt{comm}_{i,\beta})_{i,\beta}, (\texttt{comm}'_{i,\beta})_{i,\beta}, \hat{f}, \pi\right)$
14. **If** $\mathsf{NIZK}.\mathbf{Verify}\left(\mathrm{sid} = \mathsf{NIZK}_1, \left(\hat{f}, z, (\texttt{comm}_{i,\beta})_{i,\beta}, \texttt{crs}\right), \pi\right) = \texttt{reject}$
15.     **Return** $(\texttt{reject}, \perp)$ // Verify proof of garbling and relation satisfaction
16. **For** $i \in [b]$:
17.     $S_{i,0}, S_{i,1} \subset [\zeta]$ be random subsets with $|S_{i,0}| = |S_{i,1}| = \frac{\zeta}{16}$ and $S_{i,0} \cap S_{i,1} = \emptyset$
18.     **For** $\beta \in \{0,1\}$ and $\alpha \in S_{i,\beta}$: // Cut-and-choose verification of input labels
19.         $\mathsf{sh}^\alpha_{i,\beta}, \pi^\alpha_{i,\beta} \leftarrow \mathsf{OTP}.\mathbf{Eval}(\mathsf{otp}_{i,\alpha}, \beta)$
20.         **If** $\mathsf{ZK}.\mathbf{Verify}\left(\mathrm{sid} = \mathsf{NIZK}_{i,\beta}, \left(\mathsf{sh}^\alpha_{i,\beta}, \alpha, \texttt{comm}_{i,\beta}, \texttt{comm}'_{i,\beta}\right), \pi^\alpha_{i,\beta}\right) = \texttt{reject}$
21.             **Return** $(\texttt{reject}, \perp)$
22. Let $G_i = [\zeta] \setminus (S_{i,0} \cup S_{i,1})$ for $i \in [b]$ // Use remaining OTPs for evaluation
23. Let $\mathsf{otp} \leftarrow \left(\hat{f}, (G_i)_i, (\mathsf{otp}_{i,\alpha})_{i \in [b], \alpha \in G_i}, (\texttt{comm}_{i,\beta}\texttt{comm}'_{i,\beta})\right)$
24. **Return** $(\texttt{accept}, \mathsf{otp})$

Ver-$\mathsf{OTP}$.**Eval**$(\mathsf{otp}, x)$ for $x \in \{0,1\}^b$:

25. Parse $\mathsf{otp}$ as $\left(\hat{f}, (G_i)_i, (\mathsf{otp}_{i,\alpha})_{i \in [b], \alpha \in G_i}, (\texttt{comm}_{i,\beta}\texttt{comm}'_{i,\beta})\right)$
26. **For** $i \in [b], \alpha \in G_i$: // Get input label shares for input $x$
27.     $\mathsf{sh}^\alpha_{i,x_i}, \pi^\alpha_{i,x_i} \leftarrow \mathsf{OTP}.\mathbf{Eval}(\mathsf{otp}_{i,\alpha}, x_i)$
28. **For** $i \in [b]$:
29.     $\mathrm{rec}_i = \left\{\alpha \mid \alpha \in G_i, \mathsf{ZK}.\mathbf{Verify}\left(\mathrm{sid} = \mathsf{NIZK}_{i,x_i}, \left(\mathsf{sh}^\alpha_{i,x_i}, \alpha, \texttt{comm}_{i,x_i}, \texttt{comm}'_{i,x_i}\right), \pi^\alpha_{i,x_i}\right) = \texttt{accept}\right\}$
30. $\kappa_{i,x_i} \leftarrow \mathsf{Rec}\left(\left(\mathsf{sh}^\alpha_{i,x_i}\right)_{\alpha \in \mathrm{rec}_i}\right)$ // Reconstruct input label from valid shares
31. **Return** $y \leftarrow \mathbf{Eval}\left(\hat{f}, (\kappa_{i,x_i})_{i \in [b]}\right)$ // Evaluate garbled circuit to get output

**Security against malicious sender**: Let $\mathcal{A}$ be an adversary corrupting the sender. Then, we construct a simulator $\mathsf{Sim}_{\text{Ver-}\mathsf{OTP}}$ as follows for the adversary $\mathcal{A}$ by simply having the simulator execute the input given by $\mathcal{A}$. This execution determines the sender's input $s_{\mathsf{otp}}$ as well as proof $\pi$ and auxiliary information $z$. We now show that the verification step of the protocol ensures that either $\mathcal{A}$ is caught cheating or that the sender is honest. First, note that if $\pi$ does not verify, then the receiver will reject by the security of $\mathcal{F}_{\mathsf{NIZK\text{-}Arg}}$. Then, we have that if $\pi$ verifies:

- $\hat{f}$ is a valid garbling of $f(s_{\mathsf{otp}}, \cdot)$ with input wires $(\kappa_{i,\beta})_{i,\beta}$,
- $\mathtt{comm}_{i,\beta}$ are valid commitments to $\kappa_{i,\beta}$
- $\mathcal{R}(s_{\mathsf{otp}}, z) = \mathtt{accept}$

Now, we just have to show that the cut-and-choose strategy ensures that the receiver can reconstruct the garbled input wire labels for all of the input wires with overwhelming probability. By the security of $\mathcal{F}_{\mathsf{NIZK\text{-}Arg}}, \mathcal{F}_{\mathsf{Comm}}$, we have that each one-time program $\mathsf{otp}_{i,\alpha}$ and output $\mathsf{sh}^{\alpha}_{i,\beta}, \pi^{\alpha}_{i,\beta}$ for $\beta \in \{0,1\}$ either

- $\pi^{\alpha}_{i,\beta}$ verifies
- or the receiver rejects.

If the receiver rejects, then we are done. Otherwise, we have that randomly chosen subsets $S_{i,0}, S_{i,1} \subset [\zeta]$ with $|S_{i,0}| = |S_{i,1}| = \frac{\zeta}{16}$ and $S_{i,0} \cap S_{i,1} = \emptyset$ are checked. Note that in order for reconstruction of $\kappa_{i,\beta}$ to fail, then more than $\frac{\zeta}{8}$ of the $\zeta$ one-time programs for input $\beta$ must be corrupted.[8] But then, the probability that $S_{i,\beta}$ contains no corrupted one-time programs is at most $O(\exp(-\zeta))$ by a Chernoff bound as $|S_{i,\beta}| = \frac{\zeta}{16}$ and so the probability that $S_{i,\beta}$ contains at least one corrupted one-time program is approximately $1 - \left(1 - \frac{1}{8}\right)^{\frac{\zeta}{16}}$ if at least $\frac{\zeta}{8}$ of the one-time programs are corrupted. Thus, with all but exponentially small probability, the receiver can ensure that it can reconstruct $\kappa_{i,\beta}$ for $\beta \in \{0,1\}$ and thus evaluate the garbled circuit to obtain $f(s_{\mathsf{otp}}, x)$ for any input $x$ in the verification step.

Finally, we note that in the execution step, if $\mathsf{Sim}_{\text{Ver-}\mathsf{OTP}}$ outputted abort in the verification step, then we are done. Otherwise, for input $x \in \{0,1\}^b$, the receiver can reconstruct $\kappa_{i,x_i}$ for $i \in [b]$ and thus evaluate the garbled circuit to obtain $f(s_{\mathsf{otp}}, x)$ as we know that:

- $\hat{f}$ is a valid garbling of $f(s_{\mathsf{otp}}, \cdot)$ with input wire labels $(\kappa_{i,\beta})_{i,\beta}$,
- $(z, s_{\mathsf{otp}}) \sim \mathcal{R}$

**Security against malicious receiver**: Let $\mathcal{A}$ be an adversary corrupting the receiver. Then, we construct a simulator $\mathsf{Sim}_{\text{Ver-}\mathsf{OTP}}$ for adversary $\mathcal{A}$ in Simulator 5.

We now show that the output of $\mathsf{Sim}_{\text{Ver-}\mathsf{OTP}}$ is computationally indistinguishable from the real execution of the protocol. First, note that by the security of $\mathcal{F}_{\mathsf{OTP}^*}$, $\mathcal{A}$ can only evaluate each one-time program $\mathsf{otp}_{i,\alpha}$ once for $i \in [b], \alpha \in [\zeta]$ and thus for each $i \in [b]$, $(\mathsf{otp}_{i,\alpha})_{\alpha \in [\zeta]}$ can only be evaluated on more than half of its inputs for at most one of $\beta \in \{0,1\}$. And so, by the security of our secret sharing scheme (Definition 2.5), $\kappa_{i,1-x_i}$ for $i \in [b]$ is indistinguishable from random and so can be set by Simulator 5. Moreover, note that $r_{i,\beta}$ is uniformly random in the real world and so can also be set by Simulator 5. Then, as $\kappa_{i,1-x_i}$ is removed from the adversary's view, we can use the selective security of garbled circuits (Definition 2.3) to replace the garbled circuit and its output with a simulated garbled circuit and output. Then, as the proof $\hat{\pi}$ is no longer tied to the real garbled

[8] These parameters can be adjusted as long as a Chernoff-type bound can be applied. We choose these parameters for simplicity and clarity.

circuit, we will use the trapdoor $\tau$ to generate a valid proof for the simulated garbled circuit and the commitments to $\kappa'_{i,\beta}$ for $\beta \in \{0,1\}$. Finally, we note that after replacing $\mathsf{otp}_{i,\alpha}$ with updated versions using our new commitments and wire labels, we have removed all dependence on the input circuit $f(\mathtt{sk}_{\mathsf{otp}}, \cdot)$ except for $f(\mathtt{sk}_{\mathsf{otp}}, x)$ as desired.

---

**Simulator 5:** Simulator, $\mathsf{Sim}^{1\text{-}f(\mathtt{sk}_{\mathsf{otp}},\cdot)}_{\text{Ver-OTP}}(\tau, \mathcal{R}, z, \mathtt{aux})$ for verifiable one-time program with program $f(\mathtt{sk}_{\mathsf{otp}}, \cdot)$, relation $\mathcal{R}$ and trapdoor $\tau$ 1-$f(\mathtt{sk}_{\mathsf{otp}}, \cdot)$ denotes a single oracular call to $f(\mathtt{sk}_{\mathsf{otp}}, \cdot)$.

---

1 Choose $x \in \{0,1\}^b$ as follows:

2 **If** $\mathcal{A}$ makes more than half of its evaluations of $\mathsf{otp}_{i,\alpha}$ for $\alpha \in [\zeta]$ on input $\beta$ for some $\beta \in \{0,1\}$, then set $x_i = \beta$.

3 Otherwise, set $x_i$ uniformly at random in $\{0,1\}$.

4 Let $r'_{i,\beta}$ be uniformly random for $i \in [b], \beta \in \{0,1\}$ and let $\mathtt{comm}'_{i,\beta}$ be commitments to these uniformly random strings.

5 Call the one-time program to get $y = f(\mathtt{sk}_{\mathsf{otp}}, x)$

6 Set $\kappa_{i,1-x_i}$ to uniformly random for $i \in [b]$

7 Let $\mathtt{comm}'_{i,x_i}$ be a commitment to $\kappa'_{i,1-x_i}$.

8 Set $\left(\hat{f}'\left(\kappa'_{i,x_i}\right)_i\right) \leftarrow \mathsf{Garble.Sim}\left(f\left(\mathtt{sk}_{\mathsf{otp}}, \cdot\right), x, f\left(\mathtt{sk}_{\mathsf{otp}}, x\right)\right)$

9 Let $\hat{\pi}$ be a simulated proof for $\hat{\mathcal{R}}$ relative to $\left(\hat{f}', z, \left(\mathtt{comm}_{i,\beta}\right)_{i\in[b],\beta\in\{0,1\}}, \mathtt{crs}\right)$ using the trapdoor $\tau$.

10 Let $\mathtt{comm}'_{i,x_i}$ be a commitment to $\kappa'_{i,x_i}$.

11 Use the simulator for $\mathsf{OTP}^*$ to simulate the OTPs $\mathsf{sh}^{\alpha}_{i,x_i}, \pi^{\alpha}_{i,x_i} \leftarrow \mathsf{otp}_{i,\alpha}(x_i)$ for $\alpha \in [\zeta]$ and reconstruct $\kappa'_{i,x_i}$ for $i \in [b]$.

12 **Output** the simulated view to $\mathcal{A}$ consisting of:

$$\left(\left(\mathsf{otp}_{i,\alpha}\right)_{i\in[b],\alpha\in[\zeta]}, \left(\mathtt{comm}_{i,\beta}\right)_{i,\beta}, \left(\mathtt{comm}'_{i,\beta}\right)_{i,\beta}, \hat{f}', \hat{\pi}\right)$$

---

■

# 4 Multi-Key Homomorphic Encryption

Before we proceed to define open secure computation, we need to first introduce an additional primitive: multi-key homomorphic encryption (multi-key HE, or MHE).

We will use similar notation and definitions as introduced in Ananth et al. in their work on multi-key FHE in the plain model [1, 23]. We will make four main modifications, specifically:

1. We do not require the index of the party to be known ahead computation time
2. We add in support for auxiliary input in the security definition which is independent of the parties' keys
3. If not all partial decryptions are known, we allow for the partial decryptions to be simulated without knowledge of any function output.
4. We allow for computation and partial decryption to be done on *subsets* of ciphertexts and keys, rather than all of them.

We give a brief justification for why we believe these modifications are achievable with existing schemes though we leave the formalization and construction to future work.

**Modification 1**: In many applications of MHE, such as in our construction of open secure computation, the index of the party does not need to be known when encrypting the ciphertexts [1, 26].

**Modification 2**: Auxiliary input which is independent of the secret keys ensures that each LWE cipher-text is still indistinguishable from random as the secret keys are still uniformly random and independent of the auxiliary input.

**Modification 3**: An MHE cipher-text typically consists of an LWE encryption $b = \sum_i a_i \cdot s_i + e + m$ where $a_i$ are public and $s_i$ are the secret keys of the parties. Then, each partial decryption is typically of the form $p_i = b - a_i \cdot s_i$. one of the $a_i \cdot s_i$ terms is missing, then $m$ is still masked by a remaning $a_i \cdot s_i + e$ term which is indistinguishable from random by the security of LWE.

**Modification 4**: Note that in all MHE schemes (by definition), a function computed on a cipher-text do not depend on cipher-texts not included in the computation. Then, if only a subset of input ciphertexts are used, then the output of the function on the inputted ciphertexts should be independent of the missing ciphertexts.

We are now ready to present our modified MHE scheme.

**Definition 4.1** *(Multi-Key FHE* (MHE)*)*: A multiparty homomorphic encryption is a tuple of algorithms $\mathsf{MHE} = (\mathbf{KeyGen}, \mathbf{Enc}, \mathbf{Eval}, \mathbf{PartDec}, \mathbf{FinDec})$ which can be defined as follows for party $P_\ell$:

- $\mathbf{KeyGen}(1^\lambda)$ On input security parameter $\lambda$, **KeyGen** outputs a pair of keys, $(\mathtt{pk}_\ell, \mathtt{sk}_\ell)$ for the $\ell$-th party
- $\mathbf{Enc}(\mathtt{pk}_\ell, x_\ell)$ On input public key $\mathtt{pk}_\ell$ of the $\ell$-th party, and a message $x_\ell$, it outputs a ciphertext $\mathtt{ct}_\ell$
- $\mathbf{Eval}\Big(C, (\mathtt{ct}_j)_{j\in[N]}\Big)$ On input the circuit $C$ of size poly in $\lambda$ and ciphertexts, **Eval** outputs the evaluated ciphertext $\widehat{\mathtt{ct}}$
- $\mathbf{PartDec}(\mathtt{sk}_\ell, \ell, \widehat{\mathtt{ct}})$ On input $\mathtt{sk}_\ell$ of the $\ell$-th party, index $\ell$, and evaluated cipher-text $\widehat{\mathtt{ct}}$, **PartDec** outputs the partial decryption $p_\ell$ of the $\ell$-th party
- $\mathbf{FinDec}\Big(C, (p_j)_j\Big)$ On input circuit $C$ and all partial decryptions $(p_j)_{j\in N}$, **FinDec** outputs a value $y \in \{0,1\}^{C.\mathsf{out}}$

We then define the correctness of MHE as follows:

**Definition 4.2** *(*MHE *Correctness)*: An MHE scheme is said to be correct if for any inputs $(x_i)_{i\in[N]}$ and circuit $C$, the following holds for all $K \subseteq [N]$

$$\Pr\left[\begin{array}{c} \forall i, (\mathtt{pk}_i, \mathtt{sk}_i) \leftarrow \mathbf{KeyGen}(1^\lambda) \\ \mathtt{ct}_i \leftarrow \mathbf{Enc}(\mathtt{pk}_i, x_i) \\ \widehat{\mathtt{ct}} \leftarrow \mathbf{Eval}\Big(C, (\mathtt{ct}_j)_{j\in K}\Big) \\ p_i \leftarrow \mathbf{PartDec}(\mathtt{sk}_i, i, \widehat{\mathtt{ct}}), \forall i \in K \\ y \leftarrow \mathbf{FinDec}\Big(C, (p_j)_{j\in K}\Big) \end{array} \middle| \; y = C\Big(x_1, ..., x_{K_{|K|}}\Big) \right] = 1.$$

**Security of** MHE

Security is defined in a simulation-based model. In the real world, the adversary is given access to all of the public information alongside the sectet randomness of the dishonest parties. Then, we model the distributed functionality of MHE as an oracle which provides partial decryptions for the honest parties. We note that the use of this oracle can be restricted depending on the use-case. Next, in the ideal world, a simulator $\mathsf{Sim}_1$ generates the honest parties' public keys and ciphertexts as well as randomness for the dishonest parties. Then, the adversary is given access to an oracle which executes a stateful simulator $\mathsf{Sim}_2$ to obtain partial decryptions for the honest parties. Moreover, the adversary is given access to an oracle which executes a stateful simulator $\mathsf{Sim}_3$ to obtain partial decryptions for the honest parties where not all honest parties' partial decryptions are known to the adversary.

**Definition 4.3** *(Reusable Semi-Malicious Security for* MHE*)*: Let $\mathcal{A}$ be a PPT adversary, $\mathsf{Sim} = (\mathsf{Sim}_1, \mathsf{Sim}_2)$ be a PPT simulator, $H \subseteq [N]$ be the set of honest parties, and the input $(x_i)_{i\in[N]}$ be the inputs of the parties. Let $\tilde{H} = [N] \setminus H$ be the set of dishonest parties. Then, the real/ideal world is then defined in Figure 3.

$\mathsf{Real}^{\mathcal{A}}\Big(1^\lambda, H, (x_i)_{i\in H}, \mathtt{aux}\Big)$:

**for** $i \in [N]$

$r_i, r_i' \leftarrow \{0,1\}^*$

$(\mathtt{pk}_i, \mathtt{sk}_i) \leftarrow \mathbf{KeyGen}(1^\lambda, i; r_i)$

$\mathtt{ct}_i \leftarrow \mathbf{Enc}(\mathtt{pk}_i, x_i; r_i')$

**endfor**

$\mathcal{A}^{\mathcal{O}(1^\lambda,\cdot,\cdot)\mathcal{O}_\perp(1^\lambda,\cdot,\cdot,\cdot)}\Big(1^\lambda, (\mathtt{pk}_i, \mathtt{ct}_i)_{i\in H}, (r_i, r_i')_{i\in\tilde{H}}, \mathtt{aux}\Big)$

**return** $\mathsf{View}_{\mathcal{A}}$

$\mathsf{Ideal}^{\mathcal{A}}\Big(1^\lambda, H, (x_i)_{i\in H}, \mathtt{aux}\Big)(k)$:

$\Big(\mathtt{st}_S, (\mathtt{pk}_i, \mathtt{ct}_i)_{i\in H}, (r_i, r_i')_{i\in\tilde{H}}\Big) \leftarrow \mathsf{Sim}_1\Big(1^\lambda, H, (x_i)_{i\in\tilde{H}}\Big)$

$\mathcal{A}_2^{\mathcal{O}'(1^\lambda,\cdot,\cdot),\mathcal{O}_\perp'(1^\lambda,\cdot,\cdot,\cdot)}\Big(1^\lambda, (\mathtt{pk}_i, \mathtt{ct}_i)_{i\in H}, (r_i, r_i')_{i\in\tilde{H}}, \mathtt{aux}\Big)$

**return** $\mathsf{View}_{\mathcal{A}}$

*with oracles*

$\mathcal{O}(1^\lambda, C, K)$: $K \subseteq H$

$\widehat{\mathtt{ct}} \leftarrow \mathbf{Eval}\Big(C, (\mathtt{ct}_j)_{j\in K}\Big)$

**for** $i \in K, p_i \leftarrow \mathbf{PartDec}(\mathtt{sk}_i, i, \widehat{\mathtt{ct}})$

**return** $(p_i)_{i\in K}$

$\mathcal{O}'(1^\lambda, C, K)$:

$\Big(\mathtt{st}_S', (p_i)_{i\in K}\Big) \leftarrow \mathsf{Sim}_2\Big(\mathtt{st}_S, C, C\big((x_i)_{i\in K}\big)\Big)$

Update $\mathtt{st}_S = \mathtt{st}_S'$

**return** $(p_i)_{i\in K}$

$\mathcal{O}_\perp(1^\lambda, K, K')$, $K' \subsetneq K, K \subseteq H$:

$\widehat{\mathtt{ct}} \leftarrow \mathbf{Eval}\Big(C, (\mathtt{ct}_j)_{j\in K}\Big)$

**for** $i \in K', p_i \leftarrow \mathbf{PartDec}(\mathtt{sk}_i, i, \widehat{\mathtt{ct}})$

**return** $(p_i)_{i\in K'}$

$\mathcal{O}_\perp'(1^\lambda, C, K, K')$

$\Big(\mathtt{st}_S', (p_i)_{i\in K'}\Big) \leftarrow \mathsf{Sim}_3(\mathtt{st}_S, K, K')$

Update $\mathtt{st}_S = \mathtt{st}_S'$

**return** $(p_i)_{i\in K'}$

Figure 3: Real/Ideal world for MHE

Notice the addition of oracles $\mathcal{O}_\perp, \mathcal{O}_\perp'$ which are not included in the original definition of Ananth et al. [1]. These oracles allow the adversary to obtain partial decryptions for a subset of honest parties $K' \subsetneq K$ on circuit $C$ without gaining all the partial decryptions.

We then require that the simulator can simulate these partial decryptions without knowledge of the function output $C\Big((x_i)_{i\in[K]}\Big)$.

An $\mathsf{MHE}$ scheme is then said to satisfy reusable semi-honest security if the following holds: there exists a PPT simulator $\mathsf{MHE.Sim} = (\mathsf{Sim}_1, \mathsf{Sim}_2, \mathsf{Sim}_3)$ such that for every PPT $\mathcal{A}$, for any sets of honest parties $H \subseteq [N]$, any PPT distinguisher $\mathcal{D}$, and any messages $(x_i)_{i\in[H]}$, the following holds:

$$\Big|\Pr\Big[\mathcal{D}\Big(1^\lambda, \mathsf{MHE.Real}\Big(1^\lambda, H, (x_i)_{i\in[N]}\Big)\Big) = 1\Big] - \Pr\Big[\mathcal{D}\Big(1^\lambda, \mathsf{MHE.Ideal}^{\mathcal{A}}\Big(1^\lambda, H, (x_i)_{i\in[N]}\Big)\Big) = 1\Big]\Big| \leq \mathsf{negl}(\lambda).$$

# 5 Open Secure Computation

We define the ideal functionality for *open secure computation* ($\mathsf{OSC}$) where a potentially unknown and unbounded set of sending parties can send inputs to a fixed and known receiving party who can then choose to compute a function on *any subset* of the sending parties' inputs. We describe the ideal functionality in Ideal Functionality 5. Note that we assume a global setup phase where a common reference string `crs` is generated and sent to all parties. Moreover, the receiving party $R$ can *adaptively* choose to partition inputs into disjoint sets depending on the prior outputs it has received. Each disjoint set can be of arbitrary size up to $k$ and the receiving party can choose to compute the function, $f$, on *any* subset of the remaining honest parties' inputs.

We then define simulation soundness of $\mathsf{OSC}$ in the standard way:

**Definition 5.1** *(Open Secure Computation* $\mathsf{OSC}$*)*: For function

$$f : \{\{0,1\}^a \cup \bot\}^k \to \{0,1\}^b,$$

we define Open Secure Computation ($\mathsf{OSC}$) as a protocol $\pi$ between potential sending parties in $\mathcal{S}^*$ and $\mathcal{R}$ such that the following ideal and real worlds are computationally indistinguishable:

- $\mathsf{Real}_{\lambda,C,\pi,\mathcal{A}}\big(1^\lambda, (x_i)_{i\in\mathcal{X}}, y\big)$ is the real world execution of protocol $\pi$ with adversary $\mathcal{A}$ on input $\big(x_{\mathcal{X}_1}, ..., x_{\mathcal{X}_{|\mathcal{X}|}}\big)$ where parties $C$ are corrupted. Let $y$ be the output of the receiving party $R$ in the real world.
- $\mathsf{Ideal}_{\lambda,C,\mathcal{F}_{\mathsf{OSC}},\,\mathsf{Sim}(\tau)}\big(1^\lambda, (x_i)_{i\in\mathcal{X}}, y\big)$ is the ideal world with ideal functionality $\mathcal{F}_{\mathsf{OSC}}$, simulator $\mathcal{S}$ on input $\big(x_{\mathcal{X}_1}, ..., x_{\mathcal{X}_{|\mathcal{X}|}}\big)$ where the parties in $C$ are corrupted, and $\mathcal{Y} \leftarrow \mathcal{F}_{\mathsf{OSC}}(x_1, ..., x_k)$ be the output of the receiving party $R$.

**Functionality $\mathcal{F}_{\mathsf{OSC}}(x_1, ..., x_k)$**

Parameterized with corrupted parties $C$ and function $f : (\{0,1\}^a \cup \bot)^k \to \{0,1\}^b$. We also assume a global setup phase where a common reference string $\mathtt{crs}$ is generated and sent to all parties and $\tau$ is a trapdoor known only to the simulator.

**Send**: Upon receiving input $(\mathtt{send}, x, \mathtt{crs})$ for $x \in \{0,1\}^a$:

- Send $x$ to the adversary $\mathcal{S}$ and receive back a bit $b$. If $b = 1$, add the index $i$ of the sending party to corruption list $C$ and set $x = \bot$.
- Add $x$ to indexed list of inputs $\mathcal{X}$ with index $i$.

**Compute**: Upon receiving input $(\mathtt{compute}, \mathtt{crs})$ from receiving party $R$:

- Send corruption list $C$ to the receiving party $R$ and size of $\mathcal{X}$.
- Let $\mathcal{L} = \{i \in [|\mathcal{X}|] : i \notin C\}$ be the set of "left-over" indices of honest sending parties.
- Let $\mathcal{Y} = ()$
- While $\mathcal{L}$ is non-empty or the receiver aborts:
    - For the $j$-th round, send $\mathcal{L}, \mathcal{Y}$ to the receiving party $R$ and receive back a bit $b$ and indexed sets $\mathcal{S} \subseteq \mathcal{L}, \mathcal{I} \subseteq [k]$ and $|\mathcal{S}| \leq k, |\mathcal{S}| = |\mathcal{I}|$. If $b = 0$, abort.
    - Then let $\mathcal{L} = \mathcal{L} \setminus \mathcal{S}$.
    - Let input $X \in (\{0,1\}^a \cup \bot)^k$ be defined as, for each $a \in [|S_j|]$:
        - Let $\mathtt{idx} = \mathcal{I}[a]$ // the index to do the computation for
        - Let $x = \mathcal{X}[i]$ with $i = S[a]$ // the input for the given index
        - $X[\mathtt{idx}] = x$ if $i \notin C$
    - Let $R$ choose arbitrary values (including $\bot$) for all unset indices of $X$.
    - Compute $y_j = f(X)$
    - Set $\mathcal{Y} = (y_1, ..., y_{j-1}, y_j)$
- Send $\mathcal{Y}$ to receiving party $R$

Ideal Functionality 5: Ideal Open Secure Computation

## 5.1 Construction of $\mathsf{OSC}$

In Construction 6, we present our protocol for $\mathsf{OSC}$ from a conceptually simple combination of *verifiable one-time programs* and *multi-key homomorphic encryption*.

## 5.2 Protocol Soundness

Given that our protocol runs in a single round of communication, we will only consider static corruptions: i.e. the adversary must choose which parties to corrupt prior to the start of the protocol.

**Theorem 5.1** (Security of $\mathsf{OSC}$): Assume the existence of *verifiable one-time programs*, and a semi-honest *multi-key homomorphic encryption* scheme secure against chosen-plaintext attacks (Definition 4.3). Then, the above protocol (Construction 6) is simulation sound as per definition Definition 5.1.

**Construction 6:** OSC with program $f$, sending parties $(S_\ell)_{\ell \in \mathbb{Z}^+}$, and receiving party $R$

$\mathsf{OSC}.\mathtt{send}(x, \mathtt{crs})$ for party $S_\ell$ and global Ver-OTP common reference string $\mathtt{crs}$:

1 $(\mathtt{pk}_\ell, \mathtt{sk}_\ell) \leftarrow \mathsf{MHE}.\mathbf{KeyGen}(1^\lambda)$

2 $\mathtt{ct}_\ell \leftarrow \mathbf{KeyGen}.\mathbf{Enc}(\mathtt{pk}_\ell, x)$ // encrypt input

3 let $\mathcal{R}$ be the relation that checks:

  the ciphertext $\mathtt{ct}_\ell$ is well-formed under public key $\mathtt{pk}_\ell$

  $\mathtt{sk}_\ell$ is the secret key corresponding to $\mathtt{pk}_\ell$

4 define the function $f_\ell\left(\mathtt{sk}, \left(\mathtt{ct}_j\right)_{j \in [k]}\right)$ to do the following:

  **if** there is no $j$, such that $\mathtt{ct}_j = \mathtt{ct}_\ell$: // ensure that cipher-text is there

    **Output** $\perp$

  $\mathtt{idx} \leftarrow j$ such that $\mathtt{ct}_j = \mathtt{ct}_\ell$

  $\widehat{\mathtt{ct}} \leftarrow \mathsf{MHE}.\mathbf{Eval}\left(f, \left(\mathtt{ct}_j\right)_{j \in [k]}\right)$ // compute the function

  $p_\ell \leftarrow \mathsf{MHE}.\mathbf{PartDec}(\mathtt{sk}_\ell, \mathtt{idx}, \widehat{\mathtt{ct}})$ // get the partial decryption

  **Output** $p_\ell$

5 $\mathsf{v\text{-}otp} \leftarrow$ Ver-OTP.**create**$(\mathtt{sk}_\ell, \mathtt{ct}_\ell, \mathtt{crs})$ with relation $\mathcal{R}$ and function $f_\ell$

6 **send** $\mathsf{v\text{-}otp}, \mathtt{ct}_\ell, \mathtt{pk}_\ell$ to receiving party $r$

$\mathsf{OSC}.\mathtt{compute}\left((\mathsf{v\text{-}otp}_\ell, \mathtt{ct}_\ell, \mathtt{pk}_\ell)_{\ell \in [n^*]}\right)$:

7 $\mathcal{G} \leftarrow \{\}$ // The "good" set to compute on

8 **for** each $\mathsf{v\text{-}otp}$ received from party $S_\ell$:

  $(\mathtt{res}, \mathsf{otp}) \leftarrow$ Ver-OTP.**verify**$(\mathtt{crs}, \mathsf{v\text{-}otp}, (\mathtt{ct}_\ell, \mathtt{pk}_\ell), f_\ell, \mathcal{R}_\ell)$

  **if** $\mathtt{res} = \mathtt{accept}$: // only include input if the OTP verifies

9   $\mathcal{G} \leftarrow \mathcal{G} \cup \{\mathsf{otp}\}$

10 let $\mathcal{L} = \mathcal{G}$ be the set of "left-over" honestly generated one-time programs.

11 let $b, (\mathcal{X}, \mathcal{I})$ be chosen by $R$ given $\mathcal{L}$

12 let $\mathcal{Y} = ()$

13 **while** $\mathcal{L} \neq \emptyset$ and $b = 1$:

14   $(\mathtt{pk}_d, \mathtt{sk}_d) \leftarrow \mathsf{MHE}.\mathbf{KeyGen}(1^\lambda)$ // sample a dummy key

15   let input $\mathtt{ct}_x \in \mathrm{ct}^k$ be defined as follows:

16   **for** $a \in \left[|s_j|\right]$:

17     $\mathtt{idx} = \mathcal{I}_j[a], i = \mathcal{X}_j[a]$ // index for computation and the sending party's index

18     $\mathtt{ct}_x[\mathtt{idx}] = \mathtt{ct}_i$

19   **for** $\mathtt{idx} \in [k] \setminus \mathcal{I}_j$:

20     $\mathtt{ct}_x[\mathtt{idx}] = \mathsf{MHE}.\mathbf{Enc}(\mathtt{pk}_d, x'_{\mathtt{idx}})$ for $x'_{\mathtt{idx}}$ adaptively chosen by $R$

21   $\widehat{\mathtt{ct}} \leftarrow \mathsf{MHE}.\mathbf{Eval}\left(f, (\mathtt{ct}_x[j])_{j \in [k]}\right)$

22   **for** $i \in \mathcal{X}_j$:

23     $p_i \leftarrow \mathsf{v\text{-}otp}.\mathbf{Eval}(\mathsf{otp}_i, \mathtt{ct}_x)$ // get partial decryption for the input parties

24   **for** $i' \in [k] \setminus \mathcal{I}_j$:

25     $p_{i'} \leftarrow \mathsf{MHE}.\mathbf{PartDec}(\mathtt{sk}_d, i', \widehat{\mathtt{ct}})$ // partial decrypt for the dummy key

26   $y_j \leftarrow \mathsf{MHE}.\mathbf{FinDec}\left(f, (p_i)_{i \in [k]}\right)$ // get the final output

27   $\mathcal{Y} = (\mathcal{Y}, y_j)$, $\mathcal{L} = \mathcal{L} \setminus \mathcal{X}_j$ // update the left-over set and output set

28   update $b, (\mathcal{X}, \mathcal{I})$ chosen by $R$ given $(\mathcal{L}, \mathcal{Y})$ // split again based on prior output

29 **output** $\mathcal{Y}$

We provide a full proof in Section APPENDIX A and a proof sketch here.

*Proof sketch*: First, we note that all corrupted senders are filtered out by the receiver via the verification of the verifiable one-time programs. Then, the receiver can choose to "adaptively" partition the honestly generated one-time programs into disjoint sets and compute the function on any subset of the remaining honest parties' inputs. Because each sender's OTP is verified, we know that the one-time program will provide the correct partial decryption. So, the receiver can decrypt the output of $f$ on each partitioned set.

To simulate corrupted senders, note that the sender only sends one message without any previous interaction. Thus, we can trivially simulate the view of the corrupted senders by generating random inputs for the corrupted senders and generating honestly generated one-time programs for the honest senders.

Now, we consider the case where the *receiver is corrupted and potentially some senders.* Broadly, we use three main ideas:

1. replace the honestly generated one-time programs with their simulations
2. "extract" out which sets of inputs the receiver partitions via which calls to the one-time program simulators are made
3. use the security of multi-key homomorphic encryption to simulate the view of the receiver. If all partial decryptions are learned for a given partition, we use $\mathsf{MHE.Sim}_2$ to simulate the partial decryptions; otherwise, we use $\mathsf{MHE.Sim}_3$ when not all partial decryptions are known. We also use $\mathsf{MHE.Sim}_1$ to simulate the ciphertexts and public-keys for the honestly generated inputs.

After the above simulations, we note that the view of the adversary which is not the output on the partitioned inputs is entirely simulated. Thus, the view of the real world and ideal world are computationally indistinguishable by our extended security definition of MHE and Ver-OTP. ■

# 6 Applications of OSC and Ver-OTPs

In this section, we outline some potential applications of open secure computation. We do not provide full protocols or security proofs, but rather outline how the existing OSC protocol can be used to construct these applications.

## 6.1 Ver-OTP Applications

Though our primary motivation for construction of Ver-OTPs is the construction of OSC, we note that if we can keep our Ver-OTPs around for a while (i.e. with quantum memory for quantum based OTPs), the primitive is quite useful in its own right.

### 6.1.1 Blockchain-Assisted Fair Exchange

Fair exchange is a strange primitive which allows for two parties to *exchange* some information in a "fair manner." I.e. either both parties receive the other's information or neither party does. As shown by Ref. [25], fair exchange is impossible to achieve in the standard model of computation without a trusted third party. Fair exchange has many applications in digital goods exchange where we want to exchange a digital good for payment or another digital good.[9]

[9] We can think of fair exchange as a "two-party atomic swap" where either both parties receive the other's good or neither party does.

However, when combining a blockchain and Ver-OTPs, we can achieve fair exchange in a *strong* model where, unless both parties are honest, no exchange will occur!

Say Alice and Bob want to exchange digital goods $🖥_A$ and $🖥_B$ where Alice wants to ensure that relationship $R_B\left(🖥_B, \texttt{aux}\right) = \texttt{accept}$ and Bob wants to ensure that relationship $R_A\left(🖥_A, \texttt{aux}\right) = \texttt{accept}$ for some auxiliary information $\texttt{aux}$.

We then sketch the following protocol:

Sketch of Blockchain-Assisted Fair Exchange Protocol

**Goal**: Exchange digital goods $🖥_A$ and $🖥_B$ in a fair manner.

**Definitions**: For $Q \in \{A, B\}$, we will write $f_Q\left(🖥_Q, \pi_{\mathbb{B}}\right) \to 🖥_Q \cup \bot$ where $\pi_{\mathbb{B}}$ is a proof that *both* Alice and Bob have posted "accept" to blockchain $\mathbb{B}$. $f_Q\left(🖥_Q, \pi_{\mathbb{B}}\right)$:
- outputs $🖥_Q$ if $\pi_{\mathbb{B}}$ is a valid proof of inclusion on blockchain $\mathbb{B}$
- outputs $\bot$ otherwise

**Protocol**:
- **Generation**: Both Alice and Bob generate Ver-OTPs, $\textsf{v-otp}_A, \textsf{v-otp}_B$, for $f_A$ and $f_B$ with relations $R_A$ and $R_B$ respectively. They then exchange the Ver-OTPs.
- **Execution**: Alice checks that $\textsf{v-otp}_B$ is valid and then posts $\texttt{accept}$ to the blockchain $\mathbb{B}$ if $R_A\left(🖥_A, \texttt{aux}\right) = \texttt{accept}$. Bob then does the same with $\textsf{v-otp}_A$ and $R_A$. If both parties posted $\texttt{accept}$, they can then generate proofs of inclusion on the blockchain, $\pi_{\mathbb{B}}$ and evaluate their Ver-OTPs to get $🖥_A$ and $🖥_B$ for Bob and Alice respectively.

We make use of the blockchain as a coordination mechanism to ensure that both parties post $\texttt{accept}$ before either party can receive the other's good.

We then note that the fair exchange occurs simply because if one party is dishonest, then either they do not post $\texttt{accept}$, in which case neither party receives the other's good, or they send a invalid Ver-OTP in which case the honest party will not post accept and again neither party receives the other's good. In the case that a dishonest party posts accept without checking their Ver-OTP and honestly sends a valid Ver-OTP, then either the other party is dishonest, in which case there are no guarantees, or the other party is honest, in which case it still can receive the good.

### 6.1.2 Other Ver-OTP Applications

We believe that Ver-OTPs can be used to construct many other interesting applications, such as:
- *Pay-to-Run*: Ver-OTPs trivially allow for software where neither party needs to trust the other and one party can "lease" the software to another party for a one-time use.
- *Software Licensing*: Ver-OTPs can be used to construct software licenses where a user can only use the software a limited number of times (e.g. $N$ times) and the software vendor does not need to trust the user to not copy the software.
- *As a building block for other protocols*: we have already shown how Ver-OTPs can be used to construct OSC and we believe that they can be used to construct many other interesting protocols as well. We leave further exploration of this idea to future work.

## 6.2 OSC Applications

For many of these applications, it will be useful to have a registration phase for $k$ parties where the public key of the parties are on a public bulletin board, $\mathcal{B}\mathcal{b}$, with some way to verify a key's inclusion on the board (e.g. a Merkle tree proof for a Blockchain's state). We will denote a proof of the inclusion of public key $\texttt{pk}_i$ on the board as $\pi_{\mathcal{B}\mathcal{b},i}$.

### 6.2.1 Honest Majority, Single-Round Atomic Propose

We term an "atomic-propose" to be a primitive where a leader proposes a value and each party can attest to the value. We note that an "atomic-propose" is a weaker primitive than consensus as it does not require agreement on the value, only an attestation of the value. Still, atomic-proposes are the building blocks of many consensus protocols, such as Ethereum and various byzantine fault-tolerant (BFT) consensus protocols [8, 12, 33].[10]

Say that we have $N$ parties, $P_1, ..., P_N$ with associated signing public keys $\texttt{pk}_1, ..., \texttt{pk}_N$ on a public bulletin board, $\mathcal{B}\mathcal{b}$. Each public key will have an associated secret key, $\texttt{sk}_i$ which is only known to party $P_i$. We will assume that a leader party $P_L$ is chosen ahead of time (e.g. via a VRF or some other random beacon).

Sketch of Single-Round Atomic Propose Protocol

**Goal**: The leader will propose a value $v \in \{0, 1\}^a$ and output a set of signatures from a majority of the parties on the value $v$ for an honest majority.

**Definitions**: We will write $\Sigma_i$ to denote the tuple $\left(\texttt{sk}_i, \pi_{\mathcal{B}\mathcal{b},i}\right)$ which denotes the party's signing key and proof of inclusion on the bulletin board. Let $f\left(v \times (\Sigma_i)_i\right) \to v \times (\sigma_i)_i$ be the following function:

- If $> \frac{1}{2}$ of the $\Sigma_i$ inputs are "honest", output $v$ along with signatures $\sigma_i = \mathrm{Sign}(\texttt{sk}_i, v)$. We define "honest" as:
  - The input is not $\perp$
  - The input has a valid secret key for the public key, $\texttt{pk}_i$, which is on the bulletin board as proved by $\pi_{\mathcal{B}\mathcal{b},i}$
- Else, output $\perp$.

**Protocol**:

- The leader $P_L$ is the receiver in OSC protocol and sender of input $v$ to the OSC function $f$.
- Each party, $P_i$ (including the leader), is a sender with input $\Sigma_i$ to the OSC function $f$.
- The leader waits to receive the inputs from all parties (or until a timeout) and then evaluates $f$ to get attestation signatures from an honest majority.

Note that in OSC, the evaluator can *partition* the parties into different computation groups. But, as we need an honest majority, the leader cannot create more than one partition which will contain an honest majority and thus output signatures from the honest parties.

[10] An atomic-propose can be turned into BFT with an extra round of communication where each party broadcasts the attestation they received from the leader.

### 6.2.2 Single-Round Sealed Bid Auction Protocol

Again, we will assume that participating parties have registered their public keys on a public bulletin board, $\mathcal{B}\ell$. Moreover, we will assume an honest majority of sending parties and an auctioneer which is guaranteed to post the output of the protocol.[11]

Sketch of Single-Round Sealed Bid Auction Protocol

**Goal**: Pre-registered bidders will submit sealed bids to an auctioneer who will then announce the winner, their price, and a signed transaction receipt allowing the auctioneer to claim the payment from the winner.

**Definitions**: We will write $\Sigma_i$ to denote the tuple $\left(\mathtt{sk}_i, \pi_{\mathcal{B}\ell,i}, b_i\right)$ which denotes the party's signing key, proof of inclusion on the bulletin board, and bid $b_i \in \{0,1\}^a$. Let $f\left(\left(\Sigma_i\right)_i\right) \to v \times \left(\sigma_i\right)_i$ be the following function:

- If $> \frac{1}{2}$ of the $\Sigma_i$ inputs are "honest", output the highest bid $b_j$ along with
$$\sigma = \mathrm{Sign}\left(\mathtt{sk}_j, \left(b_j, \text{Pay auctioneer } b_j\right)\right)$$
where $j = \arg\max_i b_i$. We define "honest" as:
    - The input is not $\bot$
    - The input has a valid secret key for the public key, $\mathtt{pk}_i$, which is on the bulletin board as proved by $\pi_{\mathcal{B}\ell,i}$
- Else, output $\bot$.

**Protocol**:

- The auctioneer is the receiver in OSC protocol with function $f$.
- The bidders send their bids, $\Sigma_i$ to the auctioneer via the OSC protocol.
- The auctioneer waits to receive the inputs from all parties and then evaluates $f$

Again, as we require an honest majority in the function $f$, the auctioneer cannot partition the parties into multiple groups which can each contain an honest majority. And so, only one partition in the OSC protocol will contain the honest parties and thus the highest bid will be from an honest party.

Assuming that the auctioneer posts the output of the protocol, then we can see that only the highest bidder will reveal any information about their bid as $f$ only outputs the highest bid and a signature from the highest bidder. Moreover, the auctioneer is then guaranteed to receive a payout as they receive a signed transaction receipt from the highest bidder.

Second-price auctions can be done similarly by modifying the function $f$ to output the second highest bid as well.

### 6.2.3 Differentially Private Statistical Aggregation

Here, we will not assume any pre-registration of parties. Rather, we can have the parties be completely unknown ahead of time and simply send a single message to an evaluator who will then output a differentially private statistic on the inputs of the parties.

[11]This can be done via a slashing mechanism or some other economic incentive. Otherwise, we note that if the auctioneer wants to claim their payment, we can coordinate the bidder receiving their item via a smart contract which guarantees fair exchange.

Sketch of Differentially Private Statistical Aggregation Protocol

**Goal**: Unknown parties will send inputs, $x_1, ..., x_k$, to an evaluator who will then output a differentially private statistic, $g(x_1, ..., x_k)$, on the inputs of the parties.

**Definitions**: We will write $\Sigma_i$ to denote the tuple $(s_i, x_i)$ which denotes the party's secret randomness and input $x_i \in \{0,1\}^a$. Let $f\big((\Sigma_i)_i\big) \to v \times (\sigma_i)_i$ be the following function:

- Let $s = \sum s_i$ and $e = \text{NoisePRG}(s)$ where $e$ is sampled from a distribution which guarantees differential privacy (e.g. Laplace or Gaussian).
- Let $y = g(x_1, ..., x_k)$.
- Output $y + e$.

**Protocol**:

- Each party samples uniformly random $s_i$ and sends $\Sigma_i = (s_i, x_i)$ to the evaluator via the OSC protocol.
- The evaluator is the receiver in OSC protocol with function $f$.
- The evaluator waits to receive inputs and chooses when to evaluate $f$

Note that in this protocol, as long as each partition has a single uniformly random $s_i$, then the noise $e$ will be honestly sampled from the desired distribution. Thus, the output statistic for each partition will be differentially private.

### 6.2.4 Other Potential Applications

We believe that OSC can be used to construct many other interesting applications. For example:

- *Sealed-bid Voting*: OSC can be used to construct protocols where voters do not need to register ahead of time, though with weaker security guarantees (i.e. the auctioneer can *partition* the voters into multiple groups if they are fully malicious).
- *Secure Lotteries*: OSC can be used to construct secure lotteries where players do not need to register ahead of time, though again with weaker security guarantees.
- *Distributed systems compilation*: Assuming pre-registration, OSC can be used to "compile down" existing multi-round distributed protocols into a signle-round protocol where we have one party (the evaluator) run the entire protocol (who could still abort if they are malicious). We leave further exploration of this idea to future work. This can be useful in settings where we have a *cloud* party which is assumed to be reliable but not trusted with the inputs of the other parties.
- And more? OSC seems to be quite a powerful primitive and we believe that there are many more applications which can be constructed.

# 7 Conclusion

In this work, we introduce two novel primitives: verifiable one-time programs (Ver-OTPs) and open secure computation (OSCs) with the end-goal of new and useful applications of single-qubit cryptography.

The core insight is that generic one-time programs can be made verifiable via a cut-and-choose technique and that verifiable one-time programs can be used to construct single-round open secure computation. In turn, single-round open secure computation can be used to construct a variety of useful cryptographic tasks such as single-round sealed-bid auctions, atomic proposes, and (differentially private) statistical aggregation without pre-registration.

Moreover, from a theoretical perspective, verifiable one-time programs are an interesting primitive in their own right, and we believe that they may find applications beyond pay-to-use cryptography, fair exchange, and open secure computation.

There are several open questions that we leave for future work:

- Can we remove the need for a CRS in our construction? The CRS does not directly affect the protocol itself but rather is a side of effect of our security proof. The author is uncertain if this is possible, but it would be interesting to see if a CRS-free construction is possible.
- Can we construct verifiable one-time programs with better efficiency? Our construction is not particularly optimized for efficiency.
- Can we add more "fault-tolerance" to our verifiable one-time programs and OSC? Our current construction assumes an idealized, noise-free world. In the real world, one-time programs based on quantum information will be noisy and non-ideal, just as with QKD and position verification.
- Can we construct UC-secure verifiable one-time programs? Our current construction is not UC-secure and requires re-winding so that we can make use of the garbled circuit's selective security.
- Can we extend both verifiable one-time programs and OSC to the quantum computational setting? I.e. can we have quantum inputs, outputs, and/or quantum functionalities? We believe that this is possible, but leave it for future work.
- Can we either construct a multi-key HE scheme which satisfies our stronger security definition or, preferably, can we modify our construction to use a standard multi-key HE scheme?
- Finally, we ask whether there are more applications of OSC which make use of its "open" nature. Currently, only the differentially private statistical aggregation application makes use of the open nature of OSC, as it allows parties to contribute data without pre-registration. We thus wonder if there are any more applications in which the partitioning attacks are not an issue, and thus the open nature of OSC can be fully utilized.

# Acknowledgements

The author is grateful to the helpful discussions and feedback from Fabrizio Romano Genovese, Stefano Gogioso, and Matthew Coudron The author also acknowledges funding and support from NeverLocal Ltd, Neon Tetra LLC, and from the NSF Graduate Research Fellowship Program.

**AI Usage:** the author would like to acknowledge the use of language models, Gemini and Claude, in generating the SVG in Figure 1 as well as proofreading for grammar, spelling, and clarity. Moreover, the author acknowledges the use of LLMs to generate the name "atomic propose" for the primitive in Section 6.

# APPENDIX A Missing Proofs for Section 5

*Proof of Theorem 5.1*: The proof proceeds in two separate cases, one for when some sending parties are corrupted and one for when the receiving party is corrupted (alongside any number of sending parties).

**Case 1: Some sending parties are corrupted**: Let $C' \subseteq \mathcal{S}$ be the set of corrupted sending parties without the receiving party $R$ being corrupted. Note that the view of each sending party is its own input and the public key it generates and thus each corrupted party's view is trivially identical in both the real and ideal worlds.

Then, we must show that the view of the receiving party $R$ is indistinguishable in both worlds even with corrupted sending parties. We say that a pair $\texttt{ct}_i, \textsf{v-otp}_i$ is valid if $\textsf{v-otp}_i$ computes the correct partial decryption $p_i$ corresponding to ciphertext $\texttt{ct}_i$ and function $f_i$. We note that a corrupted sending party $p_i$ can either send a valid pair or an invalid one. If the pair is valid, then we note that the output of $R$ is unaffected by the party's corruption as the verifiable one-time program will output the correct partial decryption $p_i$ corresponding to $\texttt{ct}_i$. If the pair is invalid, then we note that the receiving party $R$ will reject the verifiable one-time program and set the party's input to $\perp$. Next, note that the honest parties send their inputs as valid pairs and thus each sent one-time program allows for a partial decryption of the ciphertext corresponding to the honest party's input. Because only valid pairs are used in each of the MHE computations, the MHE evaluation and final decryption will output the correct function output, $f(X)$, for each partition, and the $\textsf{v-otp}$'s will output the correct partial decryptions for each valid pair input. Thus, $R$ can fully decrypt the output of each partition correctly.

We then have that the real and ideal world are indistinguishable for $R$ as in both worlds, the corrupted party's input is either a valid input (if it sent a valid pair) or $\perp$ (if the party got corrupted).

**Case 2: The receiving party is corrupted**: Assume that the receiving party $R$ is corrupted alongside a potential set of sending parties $C' \subseteq \mathcal{S}$. We now use a set of hybrid arguments to show indistinguishability between the real and ideal worlds to build a simulator for the receiving party $R$.

Let $n^*$ be the total number of honestly generated verifiable OTPs. I.e. $|\mathcal{X}| = n^* + |C'|$.

- $\textsf{Hyb}_0$: the real world execution with adversary $\mathcal{A}$.
- $\textsf{Hyb}_{1,1}$: identical to $\textsf{Hyb}_0$ except that we replace $\textsf{v-otp}_1$ with simulator the verifiable-OTP simulator $\textsf{Sim}_{1,\ \textsf{v-otp}}$ for the first honestly generated verifiable OTP.
- $\textsf{Hyb}_{1,\ell}$ for $\ell \in \{2, ..., n^*\}$: identical to $\textsf{Hyb}_{1,\ell-1}$ except that we replace $\textsf{v-otp}_\ell$ with simulator the verifiable-OTP simulator $\textsf{Sim}_{\ell,\ \textsf{v-otp}}$ for the $\ell$-th honestly generated verifiable OTP. Formally, let

$$\texttt{aux} = \{\texttt{ct}_i, \texttt{pk}_i, \mathcal{R}_i, \textsf{v-otp}_i\}_{i \in [n^*] \setminus [\ell]} \bigcup \{\textsf{v-otp}'_i, \texttt{ct}_i, \texttt{pk}_i\}_{i \in [\ell - 1]}$$

  and then let

$$\textsf{v-otp}'_\ell \leftarrow \textsf{Sim}^{1\text{-}f_\ell(\texttt{sk}_\ell, \cdot)}_{\text{Ver-OTP}}(\tau, \mathcal{R}_\ell, (\texttt{ct}_\ell, \texttt{pk}_\ell), \texttt{aux}).$$

We provide a proof of the above hybrids in the end of this section.

- $\textsf{Hyb}_2$: We use Simulator 7 to replace the inputs to the receiving party $R$ with simulated ciphertexts and public keys generated by the MHE simulator $\textsf{MHE.Sim}_1$.

Then, we replace the partial decryptions outputted by the one-time programs with that of $\mathsf{MHE.Sim}_2$ and $\mathsf{MHE.Sim}_3$. Thus, we have that view of the adversary in $\mathsf{Hyb}_2$ is indistinguishable from the ideal functionality as the ciphertexts, public keys, and partial decryptions are all simulated. We now prove that $\mathsf{Hyb}_{1,n^*} = \mathsf{Hyb}_2$. First note that in $\mathsf{Hyb}_{1,n^*}$ the adversary can make a single call to a partial decryption for each honestly generated $\mathsf{v\text{-}otp}'_\ell$. Then, we can use the soundness of the MHE scheme to replace the ciphertexts and public keys with simulated ones using $\mathsf{MHE.Sim}_1$. We then break down the calls for each $\ell$ into two cases:

- $\ell \in \text{LookedAt}$: let $H_\ell$ be the set of input indices as defined in Simulator 7. Then, we can note that all calls to $\mathsf{v\text{-}otp}'_a$ for $a \in H_\ell$ are evaluated on the same set of honestly generated ciphertexts. Thus, we have partial decryptions for the same set of honestly generated ciphertexts and thus can use the correctness of the MHE scheme to fully decrypt the output $\widehat{\mathtt{ct}}$ of $f\Big(\left(\mathtt{ct}_i\right)_i, \left(\mathtt{ct}_o\right)_o\Big)$ for $i \in H_\ell$ and $o \in \hat{H}_\ell$ to get $y_\ell$. Then, we can use $\mathsf{MHE.Sim}_2$ to simulate the partial decryptions for all $a \in H_\ell$
- $\ell \notin \text{LookedAt}$: then, we note that $\mathsf{v\text{-}otp}'_\ell$ is evaluated on a set of honestly generated ciphertexts *without* receiving all partial decryptions $p_a$ for $a \in H_\ell$ and cipher-text $\widehat{\mathtt{ct}}_\ell = f\Big(\left(\mathtt{ct}_i\right)_i, \left(\mathtt{ct}_o\right)_o\Big)$ for $i \in H_\ell$ and $o \in \hat{H}_\ell$. Then, we can use $\mathsf{MHE.Sim}_3$ to simulate the partial decryption for $\mathsf{v\text{-}otp}'_\ell$ where we have a partial decryption for only some $a \in H_\ell$.

Thus, we have that $\mathsf{Hyb}_{1,n^*} = \mathsf{Hyb}_2$ by the correctness and security of the MHE scheme.

Finally, note that $\mathsf{Hyb}_2$ with Simulator 7 is indistinguishable from the ideal world as we have removed all input except for that of the corrupted parties and the output of the ideal functionality.

■

**Simulator 7:** Final simulator $\mathsf{Sim}$ for partition $j$ outlined in $\mathsf{Hyb}_6$ with adversary $\mathcal{A}$

**Input** The ideal functionality for the receiver and trapdoor $\tau$

1 **For** each simulated one-time program $\mathsf{v\text{-}otp}'_\ell$ with $\ell \in [n^*]$:

2 | **Let** $H_\ell$ be the ordered set of indices for input-cipher texts to 1-$f_\ell(\mathtt{sk}_\ell, \cdot)$ which are honestly generated (i.e. correspond to an honestly generated $\mathtt{ct}_i$):
i.e. $H_\ell = ((\mathtt{idx}_1, \ell_1), ..., (\mathtt{idx}_m, \ell_m))$ where $\mathtt{idx}_i$ is the index in $[k]$ of the input to $f_\ell$ and $\ell_i$ is the index of the sending party who sent the honestly generated ciphertext $\mathtt{ct}_{\ell_i}$.

3 | **Let** $\hat{H}_\ell$ be the set of non-honestly generated indices (i.e. correspond to a corrupted sending party or dummy ciphertext) inputted to 1-$f_\ell(\mathtt{sk}_\ell, \cdot)$.

4 Let $p = 0$

5 Let LookedAt $= \emptyset$

6 **For each** $\ell$ not in LookedAt,

7 | **If** for all $a \in H_\ell$, $H_a = H_\ell$ (all partial decryptions for the same set of honestly generated ciphertexts are evaluated):

8 | | **Let** $p = p + 1$

9 | | **Let** $\mathcal{S}_p = (\ell_1, ..., \ell_m)$ in $H_\ell$ and $\mathcal{I}_p = (\mathtt{idx}_1, ..., \mathtt{idx}_m)$ be the corresponding indices

10 | | **Let** LookedAt $=$ LookedAt $\cup H_\ell$

11 **Let** $\mathcal{N} = [n^*] \setminus \text{LookedAt}$ // the set of honestly generated OTPs not used in any partition

12 **Replace** $\mathtt{ct}_i, \mathtt{pk}_i$ for $i \in [n^*]$ with simulated $\mathtt{ct}_i, \mathtt{pk}_i$ by $\mathsf{MHE.Sim}_1^j\left(1^\lambda, [n^*], \{x_o\}_{i \in \hat{H}_j}\right)$ where $x_o$ is input from corrupted senders. $\mathtt{st}_S$ is the outputted state by $\mathsf{MHE.Sim}_1^j$.

13 **For** $j \in [p]$:

14 | **Let** $y_j = f\left((x_i)_{i \in H_j}, (x_i)_{i \in \hat{H}_j}\right)$ be the output of function $f$ on the inputs respecting the indices in $\mathcal{I}_j$ where the $x_i$ for $i \in \hat{H}_j$ are set by the receiving party $R$.

15 | **Replace** the one-time program oracles, $\mathsf{v\text{-}otp}'_i$, to output simulated partial decryptions for each $i \in H_j$ by replacing $f_i$ with output $p_i$ where $\left(\mathtt{st}'_S, (p_i)_{i \in H}\right) \leftarrow \mathsf{MHE.Sim}_2(\mathtt{st}_S, f, y_j)$.

16 **For each** $\ell \in \mathcal{N}$:

17 | **Let** $\mathcal{S}_\ell = \{a \mid (\mathtt{idx}, a) \in H_\ell\}$

18 | **Let** $\mathcal{S}'_\ell = \{a \mid H_a = H_\ell\}$

19 | **Replace** $\mathsf{v\text{-}otp}'_\ell$ to output simulated partial decryptions by replacing $f_\ell$ with $p_\ell$ where $\left(\mathtt{st}'_S, \{p_a\}_{a \in \mathcal{S}'_\ell}\right) \leftarrow \mathsf{MHE.Sim}_3(\mathtt{st}_S, \mathcal{S}_\ell, \mathcal{S}'_\ell)$.

We now provide the first two missing hybrid proofs.

**Lemma 1.1**: $\mathsf{Hyb}_0 \overset{c}{\approx} \mathsf{Hyb}_{1,1}$

*Proof*: Note that $\mathtt{aux}$ is independent of $(\mathtt{ct}_1, \mathtt{pk}_1, \mathtt{sk}_1)$ as it is generated by other honest parties. Thus, the view of the adversary in $\mathsf{Hyb}_0$ can be characterized by $(\mathsf{v\text{-}otp}_1, \mathtt{ct}_1, \mathtt{pk}_1, \mathtt{aux})$. We can then make use of the receiver's simulator, $\mathsf{Sim}^{1\text{-}f_1(\mathtt{sk}_1,\cdot)}_{\text{Ver-}\mathsf{OTP}}(\tau, \mathcal{R}, (\mathtt{ct}_1, \mathtt{pk}_1), \mathtt{aux})$ to replace $\mathsf{v\text{-}otp}_1$ with a simulated version $\mathsf{v\text{-}otp}_{1'}$, which is computationally indistinguishable from $\mathsf{v\text{-}otp}_1$ by the security of the Ver-$\mathsf{OTP}$ scheme.■

**Lemma 1.2**: $\mathsf{Hyb}_{1,\ell-1} \overset{c}{\approx} \mathsf{Hyb}_{1,\ell}$ for $\ell \in \{2, ..., n^*\}$

*Proof*: The proof proceeds in a similar manner to that of $\mathsf{Hyb}_0 \overset{c}{\approx} \mathsf{Hyb}_{1,1}$ except that now $\mathtt{aux}$ includes the simulated one-time programs $\mathsf{v\text{-}otp}_{1'}, ..., \mathsf{v\text{-}otp}_{(\ell-1)'}$. ■